\documentclass[a4paper,12pt]{amsart}
\usepackage{amssymb}
\usepackage{amsmath}
\usepackage{ascmac}
\usepackage{comment}
\usepackage{wrapfig}
\usepackage{txfonts}%肉厚な Times 系フォント
\usepackage{color}
\usepackage[all]{xy}%xymatrix
\usepackage{extarrows}%矢印調整
\usepackage {diagbox}%斜線

\newtheorem{thm}{Theorem}[subsection]
\newtheorem{prop}[thm]{Proposition}
\newtheorem{lem}[thm]{Lemma}

\newtheorem{rem}[thm]{Remark}

\def\o#1{\overline{#1}}
\def\u#1{\underline{#1}}
\def\C{{\mathbb C}}
\def\P{{\mathbb P}}
\def\Z{{\mathbb Z}}
\def\dfrac#1#2{{\displaystyle\frac{#1}{#2}}}
\def\sfrac#1#2{{#1}/{#2}}

\def\ds{\displaystyle}
\def\prf{\noindent{\bf Proof.} }
\def\qed{$\blacksquare$}
\def\sq{$\square$}

\makeatletter
\@addtoreset{equation}{section}

\makeatother
\title{The Pad\'e interpolation method applied to 
additive difference Painlev\'e equations 
}
\author{Hidehito Nagao}
\address{Department of Arts and Science, National Institute of Technology, Akashi College, Hyogo 674-8501, Japan}
\email{nagao@akashi.ac.jp}

\begin{document}

\begin{abstract}
We study Pad\'e interpolation problems on an additive grid, related to additive difference ($d$-) Painlev\'e equations of type $E_7^{(1)}$, $E_6^{(1)}$, $D_4^{(1)}$ and $A_3^{(1)}$. By choosing suitable Pad\'e problems, we can derive time evolution equations, scalar Lax pairs of contiguous type and determinant formulae of special solutions given in terms of hypergeometric functions, for the corresponding $d$-Painlev\'e equations.
\end{abstract}

\maketitle
%\tableofcontents
\renewcommand\baselinestretch{1.2}%行間調整

%%%%%%%%%%%%%%%%%%%%%%%%%%%
\section{Introduction}\label{sec:intro}

\noindent
We begin in Sections \ref{subsec:Peq}, \ref{subsec:Lax} and \ref{subsec:sol} by introducing the background of time evolution equations, Lax forms and hypergeometric solutions to the additive difference ($d$-) Painlev\'e equation. 
Next, in Section \ref{subsec:pade method} we introduce the background of the Pad\'e method. In Section \ref{subsec:purpose} we state the purpose and organization of this paper.
 
\subsection{The background of continuous/discrete Painlev\'e equations}\label{subsec:Peq}

\noindent
Both the second order continuous and discrete Painlev\'e equation has been well studied in mathematics and physics (e.g. \cite{Conte99, Forrester10}). In the geometric approach, for each Painlev\'e equation, K. Okamoto constructed certain rational surfaces, called the ``{\it spaces of initial values}", which parametrize all the solutions \cite{Okamoto79}. Furthermore K. Takano found that the Painlev\'e equations are uniquely determined by the spaces of initial values \cite{MMT99, ST97}. Extending these works, H. Sakai 
 proposed a certain class of second order continuous/discrete Painlev\'e equations, as mentioned below.

In Sakai's theory \cite{Sakai01}, the continuous/discrete Painlev\'e equations have been classified on the basis of the spaces of initial values connected to extended affine Weyl groups. The spaces of initial values are obtained from $\P^2$ (resp. $\P^1 \times \P^1$) by blowing up at 9 (resp. 8) points. In view of the configuration of 9 (resp. 8) points in $\P^2$ (resp. $\P^1 \times \P^1$), there exist three types of discrete Painlev\'e equations in the classification: elliptic difference ($e$-), multiplicative difference ($q$-), additive difference ($d$-) and continuous (differential) types. The only $e$-Painlev\'e equation \cite{ORG01} possesses the extended affine Weyl group symmetry of type $E_8^{(1)}$ and is obtained from the most generic configuration. All the other  Painlev\'e equations are obtained from its degeneration. 
 
The second order continuous/discrete Painlev\'e equations are classified into the 22 cases\footnote{Some $q$-Painlev\'e equations, such as a second order case of the system  \cite{KNY02-1} (see also \cite{Takenawa03}), do not belong to the list of discrete Painlev\'e equations appearing in \cite{Sakai01}.} as in Figure \ref{fig:diagram-1}.
\begin{figure}[ht]
{\small
\[
\xymatrix @C=8pt@R=5pt@M=2pt{
{\rm ell.\mbox{($e$-)}} & E_8^{(1)}\ar[d] & & & & & & & A_{1,|\alpha|^2=8}^{(1)}\ar[rd] \\  
{\rm mul.}\mbox{($q$-)} & E_8^{(1)}\ar[r]\ar[d] & E_7^{(1)}\ar[r]\ar[d] & E_6^{(1)}\ar[r]\ar[d] & D_5^{(1)}\ar[r]\ar[rd] & A_4^{(1)}\ar[r]\ar[rd] & (A_2+A_1)^{(1)}\ar[r]\ar[rd]\ar[rdd] & (A_1+A_{1,|\alpha|^2=14})^{(1)}\ar[r]\ar[rd]\ar[ru] & A_1^{(1)}\ar[rd]\ar@/^23pt/[dd] & A_0^{(1)}\ar@/^23pt/[dd] \\
{\rm add.}\mbox{($d$-)}& E_8^{(1)}\ar[r] & E_7^{(1)}\ar[r] & E_6^{(1)}\ar[rr] & & \underset{(\mbox{\footnotesize$P_{\rm VI}$})}{D_4^{(1)}}\ar[r] &\underset{(\mbox{\footnotesize$P_{\rm V}$})}{A_3^{(1)}}\ar[r]
\ar[rd] &\underset{(\mbox{\footnotesize $P_{\rm III}$})}{2(A_1)^{(1)}}\ar[r]\ar[rd] & \underset{(\mbox{\footnotesize $P_{\rm III}^{D_7^{(1)}}$})}{A_{1,|\alpha|^2=4}}^{(1)}\ar[r] & \underset{(\mbox{\footnotesize$P_{\rm III}^{D_8^{(1)}}$})}{A_0^{(1)}} &&\\
&&&&&&& \underset{(\mbox{\footnotesize$P_{\rm IV}$})}{A_2^{(1)}}\ar[r] & \underset{(\mbox{\footnotesize$P_{\rm II}$})}{A_1^{(1)}}\ar[r] & \underset{(\mbox{\footnotesize$P_{\rm I}$})}{A_0^{(1)}}&&&
}
\]
}
\caption{degeneration diagram of affine Weyl group symmetries}\label{fig:diagram-1}
\end{figure}

\noindent
Here the symbol $A \to B$ represents that $B$ is obtained from $A$ by a certain limiting procedure. The $d$-Painlev\'e equation of type $D_4^{(1)}$ and its degeneration 
arise as B\"acklund (Schlesinger) transformations of the continuous Painlev\'e equations\footnote{$P_{\rm III}^{D_i^{(1)}}$ symbolizes $P_{\rm III}$ having the surface connected to the affine root system of type $D_i^{(1)}$.} ($P_{\rm II}$, $\ldots$,$P_{\rm VI}$). The symbol 
$A_{1,|\alpha|^2=l}^{(1)}$ means the root subsystem of type $A_1^{(1)}$ whose square length of roots is $l$.
 
In this paper we put an emphasis on studying the $d$-Painlev\'e equations of type $E_7^{(1)}$, $E_6^{(1)}$, $D_4^{(1)}$ and $A_3^{(1)}$ through Pad\'e interpolated problems on the additive grid as in Table \ref{tb:work} of Section \ref{subsec:pade method}. In Sections \ref{subsec:Lax} and \ref{subsec:sol} we briefly mention the background of Lax forms and hypergeometric special solutions for the $d$-Painlev\'e equations.  

\subsection{The background of Lax forms for $d$-Painlev\'e equations}\label{subsec:Lax}

\noindent
The continuous Painlev\'e equations are obtained from deformation theory of linear differential equations of $2 \times 2$ matrix type (see \cite{JM81-2}). The continuous Painlev\'e equations are given as the forms of the compatibility conditions between the linear differential equations and the corresponding differential deformation equations. 

The $d$-Painlev\'e equation of type $D_4^{(1)}$ and its degeneration can be characterized by the same linear differential equations as the corresponding continuous Painlev\'e equations and additive deformations (Schlesinger transformations). For example, some $2 \times 2$ matrix Lax pairs for types from $d$-$D_4^{(1)}$ to $d$-$A_1^{(1)}$ have been derived in \cite{GORS98}, using a Schlesinger transformation of differential equations. We call the linear differential equation the ``{\it differential Lax form}". The $d$-Painlev\'e equations of type $E_8^{(1)}$, $E_7^{(1)}$ and $E_6^{(1)}$ do not correspond to any continuous Painlev\'e equation, and the differential $2 \times 2$ matrix Lax forms of these $d$-Painlev\'e equations have been unknown. However, some $6 \times 6$, $4 \times 4$ and $3 \times 3$ matrix Lax forms have been constructed as a certain Fuchsian system of differential equations in \cite{Boalch09} for types $d$-$E_8^{(1)}$, $d$-$E_7^{(1)}$ and $d$-$E_6^{(1)}$ respectively. 

On the other hand, it is also known that all the $d$-Painlev\'e equations are obtained from a compatibility condition of a linear additive (difference) equation and its additive deformation (e.g. additive matrix type \cite{AB06}, additive scalar type \cite{KNY17,Nagao16}). For example, some 2 $\times$ 2 matrix Lax forms have been obtained utilizing moduli spaces of difference connections on $\mathbb{P}^1$ in \cite{AB06} for type $d$-$E_6^{(1)}$ and $d$-$D_4^{(1)}$.  We call the linear additive (difference) equation the ``{\it additive (difference) Lax form}".  

\subsection{The background of hypergeometric special solutions to $d$-Painlev\'e equations}\label{subsec:sol}

The continuous Painlev\'e equations, namely $P_{\rm II}$, \ldots, $P_{\rm VI}$, admit special solutions expressible in terms of various hypergeometric functions. The  coalescence cascade of the hypergeometric functions, from the Gauss hypergeometric function ${}_2F_1$ to the Airy function, corresponds to that of continuous Painlev\'e equations, from $P_{\rm VI}$ to $P_{\rm II}$ in such works as \cite{Fuchs1907, IKSY91, KNY17}. Therefore, the $d$-Painlev\'e equation of type $D_4^{(1)}$ and its degenerations admit special solutions expressed in terms of the same hypergeometric functions as the continuous Painlev\'e equations. Also, the special solutions to the $d$-Painlev\'e equations of type $E_8^{(1)}$, $E_7^{(1)}$ and $E_6^{(1)}$ have been given in terms of the generalized hypergeometric functions $
{}_kF_l$ in \cite{Kajiwara08, KNY17}. 

 Let us define the additive shifted factorials and the HGF (the generalized hypergeometric series \cite{Bailey35, GaR04}) as follows:
 \begin{equation}\label{eq:Poch}
\begin{array}{l}
(a_1, a_2, \dots, a_i)_j=(a_1)_j(a_2)_j \cdots (a_i)_j ,\quad (a_i)_j=
\displaystyle\prod_{k=0}^{j-1}(a_i+k)=\frac{\Gamma(a_i+j)}{\Gamma(a_i)},
\end{array}
\end{equation}
\begin{equation}\label{eq:HGF}
\begin{array}{l}
{}_kF_l\left(
\begin{array}{ccc}
a_1,&\dots,&a_k\\[0mm]
b_1,&\dots,&b_l
\end{array}
; x
\right)
=\displaystyle\sum_{s=0}^{\infty}\dfrac{(a_1,\ldots,a_k)_s}{(b_1,\ldots,b_l)_s}\dfrac{x^s}{s!}.
\end{array}
\end{equation}
 
Here, we consider the condition $k=l+1$. If the parameters $a_k, b_l$ satisfy the relation $a_1+\ldots +a_{l+1}+n=b_1+\ldots+b_l$, the series are called ``{\it $n$-balanced}". If the parameters satisfy the relations $1+a_1=b_1+a_2=\ldots=b_l+a_{l+1}$ and $a_2=1+\sfrac{a_1}{2}$, the series are called ``{\it very-well-poised}". 

Then, the hypergeometric solutions to the $d$-Painlev\'e equations are summarized as in Figure \ref{fig:diagram-2}.
 
\begin{figure}[ht]
 \hspace{5mm}
{\arraycolsep=0.1pt
\begin{equation}\label{eq:solad}\nonumber
\begin{array}{ccccccccccccccccccc}
{\rm add.}\mbox{($d$-)}\quad
&E^{(1)}_8&\rightarrow&E^{(1)}_7&\rightarrow&E^{(1)}_6
&&\rightarrow&&\underset{(\mbox{$P_{\rm VI}$})}{D_4^{(1)}}&\rightarrow&\underset{(\mbox{$P_{\rm V}$})}{A_3^{(1)}}&
\rightarrow
&\underset{(\mbox{$P_{\rm III}$})}{(A_1+A_1^{\prime})^{(1)}}&\rightarrow&A^{(1)}_1&\rightarrow&{{\mathbb Z}_2}\\
&&&&&&&&&&&&\searrow&&\searrow&&&\downarrow\\
&&&&&&&&&&&&&\underset{(\mbox{$P_{\rm IV}$})}{A_2^{(1)}}&\rightarrow&\underset{(\mbox{$P_{\rm II}$})}{A_1^{(1)}}\rightarrow&&\underset{(\mbox{$P_{\rm I}$})}{1}\\
\end{array}
\end{equation}
}
 
{\arraycolsep=0.1pt
\begin{equation}\label{eq:classification1}\nonumber
\begin{array}{ccccccccccccccccccc}
{\rm sol.}\quad
&\substack{{\rm V.W.P}\\{2- \rm balanced}\\{}_9F_8} \hspace{2mm}&\rightarrow&\substack{{\rm V.W.P}\\{}_7F_6} \hspace{2mm}&\rightarrow&{}_3F_2 \hspace{2mm}
&&\rightarrow&&\underset{{\rm Gauss}}{{}_2F_1}\hspace{2mm}&\rightarrow&\underset{{\rm Kummer}}{{}_1F_1}\hspace{2mm}&
\rightarrow
&{\rm Bessel}\hspace{2mm}&\rightarrow&\times&\rightarrow&\times\\
&&&&&&&&&&&&\searrow&&\searrow&&&\downarrow\\
&&&&&&&&&&&&&\substack{{\rm Hermite}\\{\rm -Weber}}
&\rightarrow&{\rm Airy}&\rightarrow&\times\\
\end{array}
\end{equation}
}
\caption{degeneration diagram of hypergeometric solutions}\label{fig:diagram-2}
\end{figure}

Here, the symbol ${\rm V.W.P}$ means ``{\it very-well-poised}". 

\begin{rem}\label{rem:trans}{\rm \bf On the transformation formulas of hypergeometric series}

The terminating ${}_7F_6$ can be rewritten as the terminating ${}_4F_3$ (e.g. \cite{Bailey35}) and the terminating Kummer (confluent hypergeometric) function ${}_1F_1$ can be also rewritten as the terminating ${}_2F_0$ (e.g. \cite{KS98}). \sq
\end{rem}

\subsection{The background of the Pad\'e method}\label{subsec:pade method}

\noindent
There exists a certain connection among Pad\'e approximation/interpolation and continuous/discrete Painlev\'e/Garnier systems\footnote{Recently it has been shown in \cite{Mano12,MT14,MT17,NY21} that Hermite-Pad\'e approximation is related to the continuous Garnier system.}. The {\it Pad\'e method} gives time evolution equations, scalar Lax pairs and determinant formulae of special solutions simultaneously, by starting from suitable problems of Pad\'e approximation (of differential grid)/interpolation (of difference grid) as in Table \ref{tb:work}. In \cite{Yamada09-1} the Pad\'e method has been applied to continuous Painlev\'e equations of type $P_{\rm VI}$, $P_{\rm V}$, $P_{\rm IV}$ and the continuous Garnier system using differential grid (i.e. Pad\'e approximation) by Y.Yamada.
 
The Pad\'e approximation is an approximation of a given function by a rational function of given order. A typical formulation is as follows.  For a given function $\psi(x)$ analytic around $x=0$, we want to find polynomials $P(x)$ and  $Q(x)$ of degree $m$ and $n \in Z_{\ge 0}$ respectively, such that
\begin{equation} \label{eq:bibun_pade_condition}
\psi(x)=\dfrac{P(x)}{Q(x)}+O(x^{m+n+1}) \quad \mbox{(differential  grid)}.
\end{equation} 

The Pad\'e interpolation\footnote{The Pad\'e interpolation (Cauchy 1821, Jacobi 1846) is older than the Pad\'e approximation (Pad\'e 1892).} is a discrete analog of the Pad\'e approximation as follows.
 For a given sequence $\psi_s$, we want to find  polynomials $P(x)$ and $Q(x)$ of degree $m$ and $n \in Z_{\ge 0}$, by the  interpolation condition
\begin{equation} \label{eq:pade3q}
\psi_s=\frac{P (x_s)}{Q (x_s)} \quad (s=0, 1, \dots, m+n) \quad \mbox{(difference grid)}.
\end{equation}
Representative choices of the interpolating points (difference grids) are the following.

\begin{table}[htbp]
\begin{center}
\begin{tabular}{ll}
 Difference grid & Interpolating point  \\
\hline
additive & $x_s=s$ \\
additive quadratic  & $x_s=s^2+\nu s$ \\
$q$- & $x_s=q^s$ \\
$q$-quadratic & $x_s=q^s+\kappa q^{-s}$ \\
elliptic & $\ds x_s=\frac{[s+\alpha][s+\beta]}{[s+\gamma][s+\delta]}$, \\
 & $\alpha+\beta=\gamma+\delta$, \\
 & $[x]$ is the {\it theta function}
\end{tabular}
\end{center} 
\caption{Interpolating grids}\label{tb:grid}
\end{table}

As is shown in Table \ref{tb:work}, the Pad\'e method has been applied to discrete Painlev\'e/Garnier systems using the various kinds of grids mentioned above. 

\begin{table}[htbp]
\begin{center}
\begin{tabular}{|c||c|c|c|c|}
\hline
 & \multicolumn{2}{|c|}{discrete} & \multicolumn{2}{|c|}{continuous}  \\
\hline
 Grid&Garnier  & Painlev\'e & Garnier & Painlev\'e \\
\hline\hline
elliptic &\cite{Yamada17}&\cite{NTY13}&&\\
\hline
$q$-quadratic &\cite{NY21}&\cite{NY21,Yamada14}&&\\
\hline
$q$- &\cite{NY18-1, NY18-2,NY21}&\cite{Ikawa13,Nagao15,Nagao17-2,NY21}&&\\
\hline
additive quadratic &&&&\\
\hline
additive &&this paper&&\\
\hline
differential &\cite{NY18-1,NY21}&\cite{Ikawa13, Nagao17-1,NY21}&\cite{NY21,Yamada09-1}&\cite{NY21,Yamada09-1}\\
\hline
\end{tabular}
\end{center} 
\caption{Previous works of Pad\'e method}\label{tb:work}
\end{table}
  
In this paper, we apply Pad\'e interpolation on the additive grid to the additive Painlev\'e equations of type $E_7^{(1)}$, $E_6^{(1)}$, $D_4^{(1)}$ and $A_3^{(1)}$. 

\begin{rem} \label{rem:key}
{\rm {\bf On the key points of the Pad\'e method}

We have two key points on the application for the Pad\'e approximation/interpolation method \cite{Ikawa13, Nagao15, Nagao17-1, Nagao17-2, NY18-1, NY18-2, NY21, NTY13, Yamada09-1, Yamada14, Yamada17}. The first point is how to choose approximated/interpolated functions (see Table \ref{tb:Ylistad} and Remark \ref{rem:choice}). The second point is to consider two linear continuous/difference three term relations (e.g. (\ref{eq:L2L3matrixad}), called ``{\it contiguity relations}") satisfied by the error terms of the Pad\'e approximation/interpolation problems. Then the error terms can be expressed in terms of special solutions of continuous/discrete Painlev\'e/Garnier systems. Therefore the continuous/difference relations (\ref{eq:L2L3matrixad}) are the main subject in the study of the Pad\'e method, and they naturally give the evolution equations, the scalar Lax pairs and the special solutions for the corresponding continuous/discrete Painlev\'e/Garnier systems. \sq
}
\end{rem}

\begin{rem} \label{rem:SOP}
 {\bf On a connection between the Pad\'e method and the theory of semiclassical orthogonal polynomials}

The connection among semiclassical orthogonal polynomials (classical orthogonal polynomials related to a suitable weight function) and Painlev\'e/Garnier systems has been demonstrated in \cite{Magnus95}. It has been shown that coefficients of three term recurrence relations, satisfied by several semiclassical orthogonal polynomials, can be expressed in terms of solutions of Painlev\'e/Garnier systems (see \cite{Clarkson13, Nakazono13, OWF11,Van07,Witte09,Witte15,WO12} for example). Thus there exists a close connection\footnote{The theory of semiclassical orthogonal polynomials give more general solutions of Painlev\'e/Garnier systems and the Pad\'e method is simpler to compute. For example, their relation was briefly proved in \cite{Yamada09-1}.} between the Pad\'e method and the theory of semiclassical orthogonal polynomials. Namely, using both approaches, we can obtain the evolution equations, the Lax pairs and the special solutions for the corresponding Painlev\'e/Garnier systems.
\sq
\end{rem}

\subsection{The purpose and the organization of this paper}\label{subsec:purpose}

\noindent
The purpose of this paper is to apply the Pad\'e interpolation method on the additive grid to type $d$-$E_7^{(1)}$, $d$-$E_6^{(1)}$, $d$-$D_4^{(1)}$ and $d$-$A_3^{(1)}$. As the main results given 
in Section \ref{sec:main results}, the following items are presented for each type.

\vspace{3mm}
{\bf(a)} Setting of the Pad\'e interpolation problem on the additive grid.

{\bf(b)} Contiguity three term relations.

{\bf(c)} The time evolution equation of the $d$-Painlev\'e equation.

{\bf(d)} The additive difference Lax form of scalar type.

{\bf(e)} Determinant formulae of hypergeometric special solutions.\\

This paper is organized as follows: In Section \ref{sec:interpolation method} we explain the Pad\'e interpolation method through the items (a)--(e) above. In Section \ref{sec:main results} we present main results for type $d$-$E_7^{(1)}$, $d$-$E_6^{(1)}$, $d$-$D_4^{(1)}$ and $d$-$A_3^{(1)}$. In Section \ref{sec:conc} we give a summary and discuss some future problems.

%%%%%%%%%%%%%%%
\section{Pad\'e interpolation method on the additive grid}\label{sec:interpolation method}

\noindent
In this section, for the additive grid case, we explain the methods for deriving the items (a)--(e) in the main results given in Section \ref{sec:main results}. 
In the item (a), the interpolated functions and interpolated sequences are given as in Table \ref{tb:Ylistad}. 
For the items (b)--(e), we partly change the $q$-grid case \cite{Nagao15} to the additive grid case. 
%%%%%%%%%%%%%%
\subsection{(a) Setting of the Pad\'e interpolation problem on the additive grid}\label{subsec:itema}

\subsubsection{Pad\'e interpolation problem on additive grid}

 Let us consider the following interpolation problem on the additive grid. 
 
For a given function $Y(x)$, we look for functions $P_m(x)$ and $Q_n(x)$ which are polynomials of degree $m$ and $n$ $\in \Z_{\geq 0}$, satisfying the interpolation condition 
\begin{equation}\label{eq:padead}
Y(s)=\sfrac{P_m (s)}{Q_n (s)} \quad (s=0,1,\ldots, m+n).
\end{equation}

We call this problem the ``{\it Pad\'e interpolation problem on the additive grid}", since the interpolation grid $s$ is an additive sequence (see Table \ref{tb:grid}). Then we call the function $Y(x)$ and the sequences $Y_s=Y(s)$ the ``{\it interpolated function}" and ``{\it interpolated sequence}", respectively (see  Table \ref{tb:Ylistad}). Correspondingly we call both the polynomials $P_m(x)$ and $Q_n(x)$ ``{\it interpolating polynomials}". The explicit expressions of the interpolating polynomials $P_m(x)$ and $Q_n(x)$
are given in the formulae (\ref{eq:Jacobiad}) (see the item (e)).

\begin{rem}\label{rem:normalizationad}
{\rm {\bf On the common normalization factor of the polynomials $P_m(x)$ and $Q_n(x)$}

The interpolation condition (\ref{eq:padead}) can not determine the common normalization factor of the interpolating polynomials $P_m(x)$ and $Q_n(x)$. However, this normalization factor is not essential for our arguments, i.e. the main results in Section \ref{sec:main results} (see Remark \ref{rem:gaugead}). \sq
}
\end{rem}

\subsubsection{The interpolated function and sequence}

Let $a_i, b_i, c$ and  $d \in \C^{\times}$ be complex parameters.  
As is given in Table \ref{tb:Ylistad}, we set up the interpolation problems (\ref{eq:padead}) by specifying the interpolated functions $Y(x)$\footnote{The given functions $Y(x)$ are interpolated by rational functions of given order. However $Y(x)$ need not be rational functions.} and the interpolated sequences $Y_s=Y(s)$. 
\begin{table}[ht]
\begin{equation}\label{eq:Ylistad}\nonumber
\begin{tabular}{|c||c|c|c|c|}
\hline
&$d$-$E_7^{(1)}$&$d$-$E_6^{(1)}$&$d$-$D_4^{(1)}$&$d$-$A_3^{(1)}$\\
\hline\hline
$\begin{array}{c}\\Y(x)\\ \\\end{array}$&$\displaystyle\prod_{i=1}^3\dfrac{\Gamma(a_i)\Gamma(x+b_i)}{\Gamma(x+a_i)\Gamma(b_i)}$&
$\displaystyle\prod_{i=1}^2\dfrac{\Gamma (a_i)\Gamma(x+b_i)}{\Gamma(x+a_i)\Gamma(b_i)}$&
$c^x\dfrac{\Gamma(a_1)\Gamma(x+b_1)}{\Gamma(x+a_1)\Gamma(b_1)}$&
$d^x\dfrac{\Gamma(x+b_1)}{\Gamma(b_1)}$\\
\hline
$\begin{array}{c}\\Y_s\\ \\ \end{array}$&$\displaystyle\prod_{i=1}^3\dfrac{(b_i)_s}{(a_i)_s}$&
$\displaystyle\prod_{i=1}^2\dfrac{(b_i)_s}{(a_i)_s}$&
$c^s\dfrac{(b_1)_s}{(a_1)_s}$&
$d^s(b_1)_s$\\
\hline
HGF&${}_4F_3$&${}_3F_2$&${}_2F_1$&${}_2F_0$\\
\hline
\end{tabular}
\end{equation}
\caption{Interpolated functions and interpolated sequences and HGFs}\label{tb:Ylistad}
\end{table}
We note that $a_1+a_2+a_3+m-(b_1+b_2+b_3+n)=0$ is a constraint for the parameters only in the case $d$-$E_7^{(1)}$. 
 In this paper determinant formulae of special solutions are expressed in terms of the HGFs ${}_rF_l$ (the abbreviation HGF means the hypergeometric function) given in Table \ref{tb:Ylistad}, namely the 
hypergeometric functions ${}_4F_3$ (\ref{eq:E7tauad}), ${}_3F_2$ (\ref{eq:E6tauad}), ${}_2F_1$ (\ref{eq:D4tauad}), and ${}_2F_0$ (\ref{eq:A3tauad}). We note that these functions are special solutions to type $d$-$E_7^{(1)}$, $d$-$E_6^{(1)}$, $d$-$D_4^{(1)}$ and $d$-$A_3^{(1)}$ in Figure \ref{fig:diagram-2} of Section \ref{subsec:sol}.

\begin{rem} \label{rem:choice}
{\rm {\bf On the choice of the interpolated functions $Y(x)$ and sequences $Y_s$}

\noindent
One may wonder how to choose the suitable interpolated functions $Y(x)$ and sequences $Y_s=Y(s)$ in Table \ref{tb:Ylistad}. However, there is no guiding principle, i.e. only heuristics, to choose the functions $Y(x)$ and the sequences $Y_s$ in the Pad\'e interpolation method, as far as we know. In this paper we succeed in choosing $Y_s$ and $Y(x)$ suitably as follows.

\begin{table}[ht]
\begin{equation}\label{eq:Ylist}\nonumber
\begin{tabular}{|c||c|c|c|c|}
\hline
&$q$-$E_7^{(1)}$&$q$-$E_6^{(1)}$&$q$-$D_5^{(1)}$&$q$-$A_4^{(1)}$\\
\hline\hline
$\begin{array}{c}\\Y(q^s)\\ \\\end{array}$&$\displaystyle\prod_{i=1}^3\dfrac{(\beta_i;q)_s}{(\alpha_i;q)_s}$&
$\displaystyle\prod_{i=1}^2\dfrac{(\beta_i;q)_s}{(\alpha_i;q)_s}$&
$\gamma^s\dfrac{(\beta_1;q)_s}{(\alpha_1;q)_s}$&
$\delta^s(\beta_1;q)_s$\\
\hline
\end{tabular}
\end{equation}
\caption{Interpolated sequences for $q$-Painlev\'e equations in \cite{Nagao15}}\label{tb:Ylist}
\end{table}
Step 1: We can choose the suitable sequences $Y_s$ of type $d$-$E_7^{(1)}$, $d$-$E_6^{(1)}$, $d$-$D_4^{(1)}$ and $d$-$A_3^{(1)}$ by
taking the replacement replacing $q=e^{\varepsilon}$, $\alpha_i=e^{a_i\varepsilon}$, $\beta_i=e^{b_i\varepsilon}$, $\alpha_i=e^{a_i\varepsilon}$, $\gamma=c$, $\delta=-\sfrac{d}{\varepsilon}$ and the limit $\varepsilon \to 0$  
for the sequences $Y(q^s)$ of type $q$-$E_7^{(1)}$, $q$-$E_6^{(1)}$, $q$-$D_5^{(1)}$ and $q$-$A_4^{(1)}$ 
in Table \ref{tb:Ylist}, respectively. Step 2: We can guess the suitable functions $Y(x)$ in response to the chosen sequences $Y_s$, respectively. We note that $\sfrac{\alpha_1\alpha_2\alpha_3q^m}{\beta_1\beta_2\beta_3q^n}=1$ is a constraint for the parameters in the case $q$-$E_7^{(1)}$. The $q$-shifted factorial is defined by $(x;q)_n=\prod_{k=0}^{n-1}(1-xq^k)$. \sq
}
\end{rem}

\subsubsection{Time evolution}

We give the parameter shift operators $T$ as in Table \ref{tb:Tlistad}. Here the operators $T$ are called the ``{\it time evolutions}", since they specify the directions of the time evolution equations for the corresponding $d$-Painlev\'e equations. 
\begin{table}[ht]
\begin{equation}\label{eq:Tlistad}\nonumber
\begin{tabular}{|c||ccc|}
\hline
&parameters&&shifted parameters\\
\hline\hline
$d$-$E_7^{(1)}$&$(a_1,a_2,a_3,b_1,b_2,b_3,m,n)$&$\mapsto$&$(a_1+1,a_2,a_3+1,b_1+1,b_2,b_3+1,m,n)$\\
\hline
$d$-$E_6^{(1)}$&$(a_1,a_2,b_1,b_2,m,n)$&$\mapsto$&$(a_1+1,a_2,b_1+1,b_2,m,n)$\\
\hline
$d$-$D_4^{(1)}$&$(a_1,b_1,c,m,n)$&$\mapsto$&$(a_1+1,b_1+1,c,m,n)$\\
\hline
$d$-$A_3^{(1)}$&$(b_1,d,m,n)$&$\mapsto$&$(b_1+1,d,m,n)$\\
\hline
\end{tabular}
\end{equation}
\caption{Directions of time evolutions}\label{tb:Tlistad}
\end{table}

We consider yet another Pad\'e problem 
$\o{Y}(s)=\sfrac{\o{P}_m (s)}{\o{Q}_n (s)} \quad (s=0,1,\ldots, m+n)$. Here, for any object $F$ the corresponding shifts are denoted by $\o{F}:=T(F)$ and $\u{F}:=T^{-1}(F)$.

%%%%%%%%%%%%%%%%%%%%%%%%
\subsection{(b) Contiguity three term relations}\label{subsec:itemb}

\subsubsection{Contiguity relations by determinant expressions}

Let us consider two linear three term relations: $L_2(x)=0$ among $y(x), y(x+1), \o{y}(x)$ and $L_3(x)=0$ among $y(x), \o{y}(x), \o{y}(x-1)$ satisfied by fundamental solutions\footnote{The set of all linear combinations of these two solutions, i.e. $y(x)=A P_m(x)+BY(x)Q_n(x)$ where $A$ and $B$ are constants are all solutions to the two linear relations $L_2(x)=0$ and $L_3(x)=0$} $y(x)=P_m(x)$, $Y(x)Q_n(x)$, where $L_2$ and $L_3$ are given as expressions 
\begin{align}\label{eq:L2L3matrixad}
L_2(x)\propto
\begin{vmatrix}\nonumber 
y(x) & y(x+1) & \o{y}(x) \\
P_m(x) & P_m(x+1) & \o{P}_m(x)\\
Y(x)Q_n(x) & Y(x+1)Q_n(x+1) & \o{Y}(x)\o{Q}_n(x)
\end{vmatrix}, \\
L_3(x)\propto
\begin{vmatrix} 
y(x) & \o{y}(x) & \o{y}(x-1) \\
P_m(x) & \o{P}_m(x) & \o{P}_m(x-1)\\
Y(x)Q_n(x) & \o{Y}(x)\o{Q}_n(x) & \o{Y}(x-1)\o{Q}_n(x-1)
\end{vmatrix}. 
\end{align}
Here the symbol $\propto$ means the direct proportion. Then we call the linear relations $L_2=0$ and $L_3=0$ the ``{\it contiguity relations}", and the contiguity relations are the main subject in our study. \\

\subsubsection{Computation method}

Let us show the method of computation of the contiguity relations $L_2=0$ and $L_3=0$.\\
We set ${\bf y}(x):=\left[\begin{array}{c}P_m(x)\\Y(x)Q_n(x)\end{array}\right]$ and define Casorati determinants $D_i(x)$ by
\begin{equation}\label{eq:Casoratiad}
\begin{array}{l}
D_1(x):=\det[{\bf y}(x),{\bf y}(x+1)],\hspace{3mm} D_2(x):=\det[{\bf y}(x),{\o{\bf y}}(x)],\hspace{3mm} D_3(x):=\det[{\bf y}(x+1),\o{{\bf y}}(x)].
\end{array}
\end{equation}
Then the expressions (\ref{eq:L2L3matrixad}) can be rewritten as follows. 
\begin{equation}\label{eq:L2L3ad}
\begin{array}{l}
L_2(x)\propto D_1(x) \o{y}(x)-D_2(x)y(x+1)+D_3(x)y(x),
\\
L_3(x)\propto \o{D}_1(x-1)y(x)+D_3(x-1) \o{y}(x)-D_2(x)\o{y}(x-1).
\end{array}
\end{equation}
Let us define basic quantities $G(x), K(x)$ and $H(x)$ (e.g. (\ref{eq:E7GKHad}) and (\ref{eq:E6GKHad})) by
\begin{equation}\label{eq:GKHad}
\begin{array}{l}
G(x):=\sfrac{Y(x+1)}{Y(x)},\quad K(x):=\sfrac{\o{Y}(x)}{Y(x)},\quad H(x):={\rm L.C.M}(G_{\rm den}(x), K_{\rm den}(x)).
\end{array}
\end{equation}
Here an abbreviation L.C.M represents the lowest common multiple, and a symbol $\mathcal{X}_{\rm den}(x)$ (resp. $\mathcal{X}_{\rm num}(x)$) means a polynomial of the denominator (resp. numerator) in a rational function $\mathcal{X}(x)$. 
For example, in the case of $d$-$E_7^{(1)}$, $G_{\rm den}(x)=\prod_{i=1}^3 (x+a_i)$, $G_{\rm num}(x)=\prod_{i=3}^3 (x+b_i)$, $K_{\rm den}(x)=(x+a_1)(x+a_3)$ and $K_{\rm num}(x)=(x+b_1)(x+b_3)$ (see eq. (\ref{eq:E7GKHad})). Substituting these quantities into the determinants (\ref{eq:Casoratiad}), we obtain the expressions
\begin{equation}\label{eq:Drelationad}
\begin{array}{l}
D_1(x)=\dfrac{Y(x)}{G_{\rm den}(x)}E_1(z), \quad E_1(x)=G_{\rm num}(x)P_{m}(x)Q_n(x+1)-G_{\rm den}(x)P_{m}(x+1)Q_n(x), 
\\[5mm]
D_2(x)=\dfrac{Y(x)}{K_{\rm den}(x)}E_2(x), \quad E_2(x)=K_{\rm num}(x)P_{m}(x)\o{Q}_n(x)-K_{\rm den}(x)\o{P}_{m}(x)Q_n(x),
\\[5mm]
D_3(x)=\dfrac{Y(x)}{H(x)}E_3(x), \quad E_3(x)=\dfrac{H(x)}{K_{\rm den}(x)}K_{\rm num}(x)P_{m}(x+1)\o{Q}_n(x)-\dfrac{H(x)}{G_{\rm den}(x)}G_{\rm num}(x)\o{P}_{m}(x)Q_n(x+1).
\end{array}
\end{equation}
Using the interpolation condition (\ref{eq:padead}) and the form of the basic quantities $G(x)$, $K(x)$ and $H(x)$ (e.g. eqs. (\ref{eq:E7GKHad}) and (\ref{eq:E6GKHad})), we can investigate positions of zeros (e.g. $x=0, 1, \ldots, m+n-1$) and degrees of the polynomials $E_i(x)$ (e.g. the polynomial $E_1(x)$ is of degree $m+n+1$ in $x$.) in the expressions (\ref{eq:Drelationad}). Then we can simply compute the determinants $D_i(x)$ (e.g. eqs. (\ref{eq:E7Dad}) and (\ref{eq:E6Dad}))  except for some factors such as $x-f$, $x-g$ and $c_i$ in $D_i(x)$, where $f$, $g$ and $c_i$ are constants depending on parameters $a_i ,b_i ,c$ but independent of $x$ (see Remark \ref{rem:genericad}). In this way we obtain the contiguity relations $L_2=0$ and $L_3=0$ (e.g. eqs. (\ref{eq:E7L2L3ad}) and (\ref{eq:E6L2L3ad})).

\begin{rem}\label{rem:gaugead}
{\rm {\bf On the gauge invariance of the product $C_0C_1$}

\noindent
Changing the common normalization factor of the interpolating polynomials $P_m(x)$ and $Q_n(x)$, we can make an $x$-independent gauge transformation of $y(x)$ in the contiguity relations $L_2=0$ and $L_3=0$. Under the $x$-independent gauge transformation of $y(x)$: $y(x)\mapsto Gy(x)$, we can change the coefficients of $\o{y}(x), y(x-1), y(x)$ and $y(x), \o{y}(x), \o{y}(x-1)$ in $L_2=0$ and $L_3=0$ (\ref{eq:L2L3ad}) as follows. 
\begin{equation}
\begin{array}{l}
(D_1(x) : D_2(x) : D_3(x))\mapsto(\sfrac{\o{G}D_1(x)}{G} : D_2(x) : D_3(x))\\
(\o{D}_1(x-1) : D_3(x-1) : D_2(x))\mapsto(\sfrac{G\o{D}_1(x-1)}{\o{G}} : D_3(x-1) : D_2(x)).
\end{array}
\end{equation}
Let us define the coefficients $C_0$ and $C_1$ in $L_2=0$ and $L_3=0$ (e.g. eqs. (\ref{eq:E7L2L3ad}) and (\ref{eq:E6L2L3ad})) as the normalization factors of the coefficients of $\o{y}(x)$ and $y(x)$, respectively. Then $C_0$ and $C_1$ change under the gauge transformation, although the product $C_0C_1$ is a gauge invariant quantity. Furthermore, $C_0$ and $C_1$ do not appear in the final form of the $d$-Painlev\'e equations (e.g. eqs. (\ref{eq:E7eqad}) and (\ref{eq:E6eqad})). \sq
}
\end{rem}

\begin{rem}\label{rem:genericad}
{\rm {\bf On two meanings of the variables $f, g$ and parameters $m, n$}

\noindent
We use $f$ and $g$ for two different meanings. The first meaning is constants (i.e. special solutions) $f$ and $g$ which are explicitly determined in terms of parameters $a_i, b_i, m$ and $n$ by the Pad\'e interpolation problem (e.g. eqs. (\ref{eq:E7Dad}), (\ref{eq:E7L2L3ad}), (\ref{eq:E7solad}), (\ref{eq:E6Dad}), (\ref{eq:E6L2L3ad}) and (\ref{eq:E6solad})). The second meaning is generic variables (i.e. generic solutions) $f$ and $g$ $\in \P^1$ apart from the Pad\'e interpolation problem (e.g. eqs. (\ref{eq:E7eqad}), (\ref{eq:E7L1L2ad}), (\ref{eq:E6eqad}) and (\ref{eq:E6L1L2ad})), namely $f$ and $g$ are unknown functions in the $d$-Painlev\'e equation. In the items (c), (d) (resp. in the items (b) and (e)) we consider $f$ and $g$ in the second meaning (resp. in the first meaning).

Similarly, we use $m$ and $n$ for two meanings. In the first meaning, $m$ and $n$ $\in \Z_{\geq 0}$ are non-negative integer parameters (e.g. eqs. (\ref{eq:E7Yad}), (\ref{eq:E7Dad}), (\ref{eq:E7L2L3ad}), (\ref{eq:E6Yad}), (\ref{eq:E6Dad}) and (\ref{eq:E6L2L3ad})). In the second meaning, $m$ and $n$ $\in \C^{\times}$ are generic complex parameters, namely $m$ and $n$ are replaced by generic complex parameters $a_0$ and $b_0$ $\in \C^{\times}$, respectively 
(e.g. eqs. (\ref{eq:E7eqad}), (\ref{eq:E7L1L2ad}), (\ref{eq:E6eqad}) and (\ref{eq:E6L1L2ad})). In the items (c) and (d) (resp. in the items (a), (b) and (e)), we consider $m$ and $n$ in the second meaning (resp. in the first meaning). Then the result of the compatibility of the contiguity relations $L_2=0$ and $L_3=0$ also holds with respect to the second meaning. \sq
}
\end{rem}

%%%%%%%%%%%%%%%%%%%%%%%%
\subsection{(c) The time evolution equations of the $d$-Painlev\'e equation}\label{subsec:itemc}

\noindent
The computation method is as follows. Let us consider generic variables $f, g \in \P^1$ and generic parameter $a_0, b_0 \in \C^{\times}$ as in the second meaning in Remark \ref{rem:genericad}. Then we can derive the $d$-Painlev\'e equation as the necessary condition for the compatibility of the contiguity relations $L_2=0$ and $L_3=0$ (e.g. eqs. (\ref{eq:E7L2L3ad}) and (\ref{eq:E6L2L3ad})). Computing the compatibility condition, we determine three variables $\u{g}, \o{f}$ and $C_0C_1$. Expressions for variables $\u{g}$ and $\o{f}$ are obtained in terms of variables $f$ and $g$. An expression for the product $C_0C_1$ is obtained in terms of variables $f, g$ and $\o{f}$ (and hence in terms of variables $f$ and $g$).

Finally, we note the following. The first and second expressions are the $d$-Painlev\'e equation (e.g. eqs. (\ref{eq:E7eqad}) and (\ref{eq:E6eqad})). The third expression is a constraint for the product $C_0C_1$ (e.g. eqs. (\ref{eq:E7C0C1ad}) and (\ref{eq:E6C0C1ad})). 

Furthermore, thanks to the method above, we also construct scalar Lax pairs and determinant formulae of hypergeometric special solutions, while simultaneously deriving these time evolution equations, as in Section \ref{subsec:itemd} and \ref{subsec:iteme}.

%%%%%%%%%%%%%%%%%%%%%%%%
\subsection{(d) The additive difference Lax form of scalar type}\label{subsec:itemd}

\subsubsection{Scalar Lax pair}

Let us consider a linear three term equations for the unknown function $y(x)$: $L_1(x)=0$ among $y(x+1), y(x), y(x-1)$ and its deformation equation $L_2(x)=0$ among $y(x), y(x+1), \o{y}(x)$, where $L_1$ and $L_2$ are given by the expressions 
\begin{equation}\label{eq:L1L2ad}
\begin{array}{l}
L_1(x)=A_1(x)y(x-1)+A_2(x)y(x)+A_3(x)y(x+1),
\\
L_2(x)=A_4(x)\o{y}(x)+A_5(x)y(x)+A_6(x)y(x-1).
\end{array}
\end{equation}  
We call the linear three term equation $L_1=0$ and its deformation equation $L_2=0$ (\ref{eq:L1L2ad}) the ``{\it scalar Lax pair}" if the compatibility condition of the linear equations $L_1=0$ and $L_2=0$ (\ref{eq:L1L2ad}) is equivalent to a $d$-Painlev\'e equation. We note that the scalar Lax pair $L_1=0$ and $L_2=0$ is equivalent to the pair of contiguity relations $L_2=0$ and $L_3=0$.\\

\subsubsection{Computation method}

Let us show how to compute the scalar Lax pair. Similarly to the item (c), we consider generic variables $f, g \in \P^1$ and generic parameters $a_0, b_0 \in \C^{\times}$ as in the second meaning in Remark \ref{rem:genericad}. We derive the Lax pair $L_1=0$ and $L_2=0$, which satisfies the compatibility condition, by using the results of the items (a)--(c) as follows. Firstly the Lax equation $L_2=0$ (e.g. eqs. (\ref{eq:E7L1L2ad}) and (\ref{eq:E6L1L2ad})) in the item (d) is the same as the contiguity relation $L_2=0$ (e.g. eqs. (\ref{eq:E7L2L3ad}) and (\ref{eq:E6L2L3ad})) in the item (b) under an $x$-independent gauge transform of $y(x)$ and changes of parameters. Secondly the Lax equation $L_1=0$ can be obtained as follows. Combining the contiguity relations $L_2=0$ and $L_3=0$ (e.g. eqs. (\ref{eq:E7L2L3ad}) and (\ref{eq:E6L2L3ad})) under generic variables $f, g$ and generic parameters $a_0, b_0$, one obtains a linear equation among the three terms $y(x+1), y(x)$ and $y(x-1)$ (see Figure \ref{fig:L1}), whose coefficient functions depend on the variables $f, g, \o{f}, C_0$ and $C_1$. 
\begin{figure}[ht]
\begin{center}\setlength{\unitlength}{1mm}
\begin{picture}(50,30)(-5,-5)
\put(-2,25){$\o{y}(x-1)$}
\put(18,25){$\o{y}(x)$}
\put(-2,-5){$y(x-1)$}
\put(0,-1){\line(1,0){43}}
\put(18,-5){$y(x)$} 
\put(39,-5){$y(x+1)$}
\put(1,21){\line(1,0){20}}
\put(21,1){\line(0,1){20}}
\put(1,21){\line(1,-1){20}}
\put(10,14){$L_3(x)$}
\put(22,0){\line(1,0){20}}
\put(22,0){\line(0,1){20}}
\put(22,20){\line(1,-1){20}}
\put(23,4){$L_2(x+1)$}
\put(0,0){\line(1,0){20}}
\put(0,0){\line(0,1){20}}
\put(0,20){\line(1,-1){20}}
\put(3,4){$L_2(x)$}
\put(-15,-5){$L_1(x)$:}
\end{picture}
\caption{Derivation of $L_1(x)$}\label{fig:L1}
\end{center}
\end{figure}

However, the variables $C_0$ and $C_1$ appear through the product $C_0C_1$. Therefore, expressing $\o{f}$ (e.g. eqs. (\ref{eq:E7eqad}), (\ref{eq:E6eqad})) and $C_0C_1$ (e.g. eqs. (\ref{eq:E7C0C1ad}) and (\ref{eq:E6C0C1ad})) only in terms of $f$ and $g$, one obtains the Lax equation $L_1=0$ (e.g. eqs. (\ref{eq:E7L1L2ad}) and (\ref{eq:E6L1L2ad})). 

Then, under generic variables $f, g$ and generic parameters $a_0, b_0$, the $d$-Painlev\'e equation (e.g. eqs. (\ref{eq:E7eqad}) and (\ref{eq:E6eqad})) is necessary and sufficient for the compatibility of the Lax pair $L_1=0$ and $L_2=0$ (e.g. eqs. (\ref{eq:E7L1L2ad}) and (\ref{eq:E6L1L2ad})). The proof for the case of $d$-$E_7^{(1)}$ will be shown in Appendix \ref{sec:Proof}. The other cases $d$-$E_6^{(1)}$, $d$-$D_4^{(1)}$ and $d$-$A_4^{(1)}$ are similarly proved. 

%%%%%%%%%%%%%%%%%%%%%%%%%%
\subsection{(e) Determinant formulae of hypergeometric special solutions}\label{subsec:iteme}

\noindent
By construction, expressions for $f$ and $g$ as in the first meaning in Remark \ref{rem:genericad} give a special solution for the $d$-Painlev\'e equation.
We present how to compute determinant formulae of the special solutions.

\subsubsection{Determinant formulae on the additive grid}

We derive the formulae (\ref{eq:Jacobiad}), which are convenient for computing the special solutions $f$ and $g$. For a given sequence $Y_s$, the polynomials $P_m(x)$ and $Q_n(x)$ of degree $m$ and $n$ for an interpolation problem 
\begin{equation} \label{eq:pade}
Y_s=\sfrac{P_m (x_s)}{Q_n (x_s)} \quad (s=0, 1, \dots, m+n)
\end{equation}
are given by the determinant expression\footnote{General Pad\'e interpolation problems have been formulated and some universal determinant
formulae for the solutions have been proposed in \cite{Noumi15}.}:
\begin{equation}\label{eq:Jacobi}
P_{m}(x)=F(x)\det\Big[\sum^{m+n} _{s=0}u_s \dfrac{x_s^{i+j}}{x-x_s }\Big]^n _{i,j=0}, \quad Q_n(x)=\det\Big[\sum^{m+n} _{s=0}u_s x_s^{i+j}(x-x_s )\Big]^{n-1} _{i,j=0},
\end{equation}
where $u_s =\sfrac{Y_s}{F^{\prime}(x_s)}$ and $F(x)=\prod_{i=0}^{m+n}(x-x_i )$.

In the additive grid case of the problem (\ref{eq:pade}) (i.e., the problem (\ref{eq:padead})), the formulae  (\ref{eq:Jacobi}) takes the  forms
\begin{equation}\label{eq:Jacobiad}
\begin{array}{l}
P_{m}(x)=\dfrac{F(x)}{(-(m+n))_{m+n}^{n+1}}\det\Big[\displaystyle\sum^{m+n} _{s=0}Y_s\dfrac{(-(m+n))_s }{s!}\dfrac{s^{i+j}}{x-s}\Big]^n _{i,j=0},
\\
Q_n(x)=\dfrac{1}{(-(m+n))_{m+n}^{n}}\det\Big[\displaystyle\sum^{m+n} _{s=0}Y_s\dfrac{(-(m+n))_s }{s!}s^{i+j}(x-s)\Big]^{n-1} _{i,j=0}.
\end{array}
\end{equation}

In the derivation of (\ref{eq:Jacobiad}), we have used the differential coefficient
\begin{align}\label{eq:Fprime}
F^{\prime}(x_s)=&(x_s -x_0 )\dots(x_s -x_{s-1})(x_s -x_{s+1})\dots(x_s -x_{m+n})\nonumber \\
=&(s-0)(s-1)\cdots \{s-(s-1)\}\{s-(s+1)\}\cdots \{s-(m+n)\}\nonumber\\
=&\sfrac{s!(-(m+n))_{m+n}}{(-(m+n))_s}.  
\end{align}

Moreover, substituting the values of $Y_s$ (\ref{eq:Ylistad}) and $F^{\prime}(x_s )$ (\ref{eq:Fprime}) into the formulae (\ref{eq:Jacobi}), one obtains the determinant formulae (\ref{eq:Jacobiad}).

\subsubsection{Computation method}

Let us demonstrate how to compute the special solutions $f$ and $g$. We can derive the expressions for the special solutions $f$ and $g$ by comparing the determinants $D_i(x)$ in eq. (\ref{eq:Drelationad}) and $D_i(x)$ (e.g. eqs. (\ref{eq:E7Dad}) and (\ref{eq:E6Dad})) in the item (b) as the identity with respect to the variable $x$ and by applying the formulae (\ref{eq:Jacobiad}). 

In case of $d$-$E_7^{(1)}$,  we make the following calculation. Firstly substituting $x=-a_i$ $(i=1, 2)$ into the determinants $D_1(x)$ in eq. (\ref{eq:Drelationad}) and $D_1(x)$ in (\ref{eq:E7Dad}) respectively, we construct an expression for the special solution $f$ in the first equation of eq. (\ref{eq:E7solad}) by comparing the two expressions for $D_1(x)$ and by applying the formulae (\ref{eq:Jacobiad}). Similarly substituting $x=-b_i$ $(i=2, 3)$ into the determinants $D_3(x)$ in eq. (\ref{eq:Drelationad}) and $D_3(x)$ in eq. (\ref{eq:E7Dad}) respectively, we construct an expression for the special solution $g$ in the second equation of eq. (\ref{eq:E7solad}) by comparing the two expressions for $D_3(x)$ and by applying the formulae (\ref{eq:Jacobiad}).

%%%%%%%%%%%%%%%%%%%%%%%%
\section{main results}\label{sec:main results}

\noindent
In this section, for each case $d$-$E_7^{(1)}$, $d$-$E_6^{(1)}$, $d$-$D_4^{(1)}$ and $d$-$A_3^{(1)}$, we show the main results by the method in Section \ref{sec:interpolation method}.

We use the notations
\begin{equation}\label{eq:notationad}
\begin{array}{l}
\sfrac{a_1a_2\dots a_n}{b_1b_2\dots b_n}:=\dfrac{a_1a_2\dots a_n}{b_1b_2\dots b_n},\quad {\mathcal N}(x):=\prod_{i=0}^{m+n-1}(x-i),\\[3mm]
T_{a_i}(F):=F |_{a_i \to a_i+1},\quad T_{a_i}^{-1}(F):= F |_{a_i \to a_i-1},
%\\ \o{F}=T(F),\quad \u{F}=T^{-1}(F),
\end{array}
\end{equation}
for any quantity (or function) $F$ depending on variables $a_i$ and $b_i$.

%%%%%%%%%%%%%%%%%%%%%%%%
\subsection{Case $d$-$E_7^{(1)}$}\label{subsec:E7ad}

\noindent
{\bf(a)} Setting of the Pad\'e interpolation problem on the additive grid

%E7ad
In Table (\ref{eq:Ylistad}) the interpolated function, the interpolated sequence and the constraint for the parameters are set up as
\begin{equation}\label{eq:E7Yad}
Y(x):=\displaystyle\prod_{i=1}^3\dfrac{\Gamma(a_i)\Gamma(x+b_i)}{\Gamma(x+a_i)\Gamma(b_i)}, \quad Y_s=\displaystyle\prod_{i=1}^3\dfrac{(b_i)_s}{(a_i)_s}, \quad m+\sum_{i=1}^3 a_i =n+\sum_{i=1}^3 b_i, 
\end{equation}
and in Table (\ref{eq:Tlistad}) the time evolution is chosen as
\begin{equation}\label{eq:E7Tad}
T: (a_1,a_2,a_3,b_1,b_2,b_3,m,n) \mapsto (a_1+1,a_2,a_3+1,b_1+1,b_2,b_3+1,m,n).
\end{equation}
\\

%E7ad
\noindent
{\bf(b)} Contiguity three term relations

By Definition (\ref{eq:GKHad}) we have the basic quantities
\begin{equation}\label{eq:E7GKHad}
\displaystyle G(x)=\prod_{i=1}^{3}\dfrac{(x+b_i)}{(x+a_i)},\quad 
K(x)=\prod_{i=1, 3}\dfrac{a_i(x+b_i)}{b_i(x+a_i)},\quad H(x)=b_1b_3\prod_{i=1}^{3}(x+a_i),
\end{equation}
and by the expression (\ref{eq:Drelationad}) we obtain the Casorati determinants
\begin{equation}\label{eq:E7Dad}
\begin{array}{l}
D_1(x)=\sfrac{c_0(x-f){\mathcal N}(x)Y(x)}{G(x)_{\rm den}},\quad
D_2(x)=\sfrac{c_1(x-h)(x-m-n){\mathcal N}(x)Y(x)}{K(x)_{\rm den}},
\\[5mm]
D_3(x)=\sfrac{c_1(x-g){\mathcal N}(x)Y(x)\prod_{i=1,3}(x+b_i)}{H(x)},
\end{array}
\end{equation}
where $h=g+a_2-b_1-b_3-n$. Here $f$, $g$, $c_0$ and $c_1$ are constants depending on parameters $a_i, b_i \in \C^{\times}(i=1,2,3)$ and $m,n \in \Z_{\geq 0}$ but independent of $x$. Then the contiguity relations $L_2=0$ and $L_3=0$ are expressed by
\begin{equation}\label{eq:E7L2L3ad}
\begin{array}{l}
L_2(x)=C_0(f-x)\o{y}(x)-(x+a_2)(x-m-n)(x-h)y(x+1)
+(x+b_1)(x+b_3)(x-g)y(x),\\[5mm]
L_3(x)=C_1(x-\o{f}-1)y(x)+(x+a_1)(x+a_3)(x-g-1)\o{y}(x)
-x(x+b_2-1)(x-h)\o{y}(x-1),
\end{array}
\end{equation}
where $C_0=\sfrac{b_1b_3c_0}{c_1}$ and $C_1=\sfrac{a_1a_3\o{c}_0}{c_1}$.

Take note that in the items (c) and (d) below we study the contiguity relations $L_2=0$ and $L_3=0$ (\ref{eq:E7L2L3ad}) for generic complex parameters $a_0, b_0$ (replacing $m, n \in \Z_{\geq 0}$ by $a_0, b_0 \in \C^{\times}$) and generic variables $f, g \in \P^1$ (depending on parameters $a_i, b_i \in \C^{\times}, i=0,1,2,3$) apart from the Pad\'e interpolation problem (\ref{eq:padead}) with eqs. (\ref{eq:E7Yad}) and (\ref{eq:E7Tad}) (see Remark \ref{rem:genericad}).\\

%E7ad
\noindent
{\bf(c)} The time evolution equations of the $d$-Painlev\'e equation

Compatibility of the contiguity relations $L_2=0$ and $L_3=0$ (\ref{eq:E7L2L3ad}) gives the evolution equations and the product $C_0C_1$ as follows.
\begin{equation}\label{eq:E7eqad}
\begin{array}{l}
\dfrac{(f-h)(f-\u{h}+1)}{(f-g)(f-\u{g})}=\dfrac{A_2(f)}{A_1(f)}, \quad \dfrac{(f-h)(\o{f}-h+1)}{(f-g)(\o{f}-g)}=\dfrac{A_2(h)}{A_1(g)}
\end{array}
\end{equation}
and
\begin{equation}\label{eq:E7C0C1ad}
\begin{array}l
C_0C_1=\sfrac{(h-g)(h-g-1)A_1(g)}{(f-g)(\o{f}-g)}\\
\phantom{C_0C_1}=\sfrac{(h-g)(h-g-1)A_2(h)}{(f-h)(\o{f}-h+1)}, 
\end{array}
\end{equation}
where $A_1(x)=(x+a_2)(x+b_2)(x+1)(x-a_0-b_0)$ and $A_2(x)=\prod_{i=1,3}(x+a_i)(x+b_i)$.

The evolution equations (\ref{eq:E7eqad}) are equivalent to the $d$-Painlev\'e equation of type $E_7^{(1)}$ given in \cite{Kajiwara08, KNY17, Nagao16, RGTT01, Sakai07}. 
The eight singular points in coordinates $(f,g)$ are on the two lines $f=g$ and $f=h(=g+a_2-b_1-b_3-b_0)$ as follows.
\begin{equation}\label{eq:E78pad}
\begin{array}{l}
(f,g)=(-a_2,-a_2), (-b_2,-b_2), (-1,-1), (-a_0-b_0,-a_0-b_0), \\
\phantom{(f,g)=}(-a_1,a_3-b_2+a_0), (-b_3, b_1-a_2+b_0), (-a_3,a_1-b_2+a_0), (-b_1, b_3-a_2+b_0).
\end{array}
\end{equation}

%E7ad
\noindent
{\bf(d)} The additive difference Lax form of scalar type

The contiguity relations $L_2=0$ and $L_3=0$ (\ref{eq:E7L2L3ad}) give two scalar additive Lax equations $L_1=0$ and $L_2=0$ expressed by
\begin{equation}\label{eq:E7L1L2ad}
\begin{array}{l}
L_1(x)=(g-h)\Big[\dfrac{A_1(g)}{(f-g)(x-g-1)}-\dfrac{A_2(h)}{(f-h)(x-h)}\Big]y(x)
\\[5mm]\phantom{L_1(x)}
-\dfrac{x\prod_{i=1}^3(x+b_i-1)}{x-f-1}\Big[y(x-1)-\dfrac{(x-h-1)(x-a_0-b_0-1)(x+a_2-1)}{(x-g-1)(x+b_1-1)(x+b_3-1)}y(x)\Big]
\\[5mm]\phantom{L_1(x)}
-\dfrac{(x-a_0-b_0)\prod_{i=1}^3(x+a_i)}{x-f}\Big[y(x+1)-\dfrac{(x-g)(x+b_1)(x+b_3)}{(x-h)(x-a_0-b_0)(x+a_2)}y(x)\Big],
\\[5mm]
L_2(x)=(f-x)\o{y}(x)-(x+a_2)(x-h)(x-a_0-b_0)y(x+1)
+(x+b_1)(x+b_3)(x-g)y(x), 
\end{array}
\end{equation}
where $h$ and $A_i(x)$ are given in the Casorati determinant  (\ref{eq:E7Dad}) and the evolution equations (\ref{eq:E7eqad}).

The additive Lax form of scalar type $L_1=0$ and $L_2=0$ (\ref{eq:E7L1L2ad}) is equivalent to the scalar ones in \cite{KNY17,  Nagao16} by using suitable gauge transformations of $y(x)$. On the other hand, the differential $4 \times 4$ matrix Lax form has been given as a certain Fuchsian system of differential equations in \cite{Boalch09}.

\noindent
%E7ad
{\bf(e)} Determinant formulae of hypergeometric special solutions

The hypergeometric solutions are constructed as the explicit forms
\begin{equation}\label{eq:E7solad}
\begin{array}{l}
\dfrac{f+a_i}{f+b_j}=\alpha\dfrac{T_{a_i}(\tau_{m,n})T_{a_i}^{-1}(\tau_{m+1,n-1})}
{T_{b_j}^{-1}(\tau_{m,n})T_{b_j}(\tau_{m+1,n-1})} \quad  (i, j=1,2,3),\quad 
\dfrac{g+a_2}{h+a_1}=\beta\dfrac{T_{a_2}(\o{\tau}_{m,n})T_{a_2}^{-1}(\tau_{m+1,n-1})}
{T_{a_1}(\tau_{m,n})T_{a_1}(\o{\tau}_{m+1,n-1})},
\end{array}
\end{equation}
where the determinant $\tau_{m,n}$ is given by
\begin{equation}\label{eq:E7tauad}
\begin{array}l
\ds \tau_{m,n}=\det\Big[(b_1)_i(a_1-j)_j{}_4F_3 \Big(\substack{\displaystyle{b_1+i, b_2, b_3 ,-(m+n)}\\[3mm]{\displaystyle{a_1-j, a_2, a_3}}},1\Big)\Big]^n _{i,j=0}, \\[5mm]
\alpha=-\dfrac{(a_i+m+n)(a_i-1)^n(b_j-1)^n}{a_i^{n+1}b_j^n}\dfrac{\prod_{k=1}^3(b_k-a_i)}{\prod_{k=1}^3(a_k-b_j)},\\[5mm]
\beta=\dfrac{b_1b_3(a_2+m+n)(a_2-1)^n(b_2-a_2)}{a_2^{n+1}a_3(b_1-a_1)(b_3-a_1)}.
\end{array}
\end{equation}
  
These determinant formulae of the generalized hypergeometric solutions (\ref{eq:E7solad}) are given in terms of the hypergeometric function ${}_4F_3$. A certain $1 \times 1$ determinant formula has been constructed in terms of the hypergeometric function ${}_7F_6$ in \cite{Kajiwara08, KNY17}. The terminating ${}_7F_6$ can be transformed into the terminating ${}_4F_3$ (see Remark \ref{rem:trans}).

%%%%%%%%%%%%%%%%%%%%%%%%
\subsection{Case $d$-$E_6^{(1)}$}\label{subsec:E6ad}

\noindent
{\bf(a)} Setting of the Pad\'e interpolation problem on the additive grid

%E6ad
In Table (\ref{eq:Ylistad}) the interpolated function and the interpolated sequence are set up as
\begin{equation}\label{eq:E6Yad}
Y(x):=\displaystyle\prod_{i=1}^2\dfrac{\Gamma(a_i)\Gamma(x+b_i)}{\Gamma(x+a_i)\Gamma(b_i)}, \quad Y_s=\displaystyle\prod_{i=1}^2\dfrac{(b_i)_s}{(a_i)_s},
\end{equation}
and in Table (\ref{eq:Tlistad}) the time evolution is chosen as
\begin{equation}\label{eq:E6Tad}
T: (a_1,a_2,b_1,b_2,m,n) \mapsto (a_1+1,a_2,b_1+1,b_2,m,n).
\end{equation}

%E6ad
\noindent
{\bf(b)} Contiguity three term relations

By Definition (\ref{eq:GKHad}) we have the basic quantities
\begin{equation}\label{eq:E6GKHad}
\displaystyle G(x)=\prod_{i=1}^{2}\dfrac{(x+b_i)}{(x+a_i)},\quad 
K(x)=\dfrac{a_1(x+b_1)}{b_1(x+a_1)},\quad H(x)=b_1\prod_{i=1}^{2}(x+a_i),
\end{equation}
and by the expression (\ref{eq:Drelationad}) we obtain the Casorati determinants
\begin{equation}\label{eq:E6Dad}
\begin{array}{l}
D_1(x)=\sfrac{c_0(x-f){\mathcal N}(x)Y(x)}{G(x)_{\rm den}},\quad
D_2(x)=\sfrac{c_1(x-m-n){\mathcal N}(x)Y(x)}{K(x)_{\rm den}},
\\[5mm]
D_3(x)=\sfrac{c_1(x-g){\mathcal N}(x)Y(x)(x+b_1)}{H(x)},
\end{array}
\end{equation}
where $f, g, c_0$ and $c_1$ are constants depending on parameters $a_i, b_i \in \C^{\times}(i=1,2)$ and $m,n \in \Z_{\geq 0}$ but independent of $x$. Then the contiguity relations $L_2=0$ and $L_3=0$ are expressed by
\begin{equation}\label{eq:E6L2L3ad}
\begin{array}{l}
L_2(x)=C_0(f-x)\o{y}(x)-(x-m-n)(x+a_2)y(x+1)+(x+b_1)(x-g)y(x),\\
L_3(x)=C_1(x-\o{f}-1)y(x)+(x+a_1)(x-g-1)\o{y}(x)
-x(x+b_2-1)(x-h)\o{y}(x-1),
\end{array}
\end{equation}
where $C_0=\sfrac{b_1c_0}{c_1}$ and $C_1=\sfrac{a_1\o{c}_0}{c_1}$.

Take note that in the items (c) and (d) below we study the contiguity relations $L_2=0$ and $L_3=0$ (\ref{eq:E6L2L3ad}) for generic complex parameters $a_0, b_0$ (replacing $m, n \in \Z_{\geq 0}$ by $a_0, b_0 \in \C^{\times}$) and generic variables $f, g \in \P^1$ (depending on parameters $a_i, b_i \in \C^{\times}, i=0,1,2$) apart from the Pad\'e interpolation problem (\ref{eq:padead}) with eqs. (\ref{eq:E6Yad}) and (\ref{eq:E6Tad}) (see Remark \ref{rem:genericad}).\\

%E6ad
\noindent
{\bf(c)} The time evolution equations of the $d$-Painlev\'e equation

Compatibility of the contiguity relations $L_2=0$ and $L_3=0$ (\ref{eq:E6L2L3ad}) gives the evolution equations and the product $C_0C_1$ as follows.
\begin{equation}\label{eq:E6eqad}
\begin{array}{l}
(f-g)(f-\u{g})=\dfrac{(f+a_2)(f+b_2)(f+1)(f-a_0-b_0)}{(f+a_1)(f+b_1)},\\
(f-g)(\o{f}-g)=\dfrac{(g+a_2)(g+b_2)(g+1)(g-a_0-b_0)}{(g-a_1+b_2-a_0)(g+a_2-b_1-b_0)}
\end{array}
\end{equation}
and
\begin{equation}\label{eq:E6C0C1ad}
\begin{array}l
C_0C_1=\sfrac{(g+a_2)(g+b_2)(g+1)(g-a_0-b_0)}{(f-g)(\o{f}-g)}\\
\phantom{C_0C_1}=(g-a_1+b_2-a_0)(g+a_2-b_1-b_0).
\end{array}
\end{equation}

The evolution equations (\ref{eq:E6eqad}) are equivalent to the $d$-Painlev\'e equation of type $E_6^{(1)}$ given in \cite{Kajiwara08, KNY17, Nagao16, RGTT01, Sakai07}. The eight singular points in coordinates $(f,g)$ are on the three lines $f=g$, $f=\infty$ and $g=\infty$ as follows.
\begin{equation}\label{eq:E68pad}
\begin{array}{l}
(f,g)=(-a_2,-a_2), (-b_2,-b_2), (-1,-1), (a_0+b_0, a_0+b_0), \\
\phantom{(f,g)=}(\infty,a_3-b_2+a_0), (\infty, b_1-a_2+b_0), (-a_1,\infty), (-b_1, \infty).
\end{array}
\end{equation}

%E6ad
\noindent
{\bf(d)} The additive difference Lax form of scalar type

The contiguity relations $L_2=0$ and $L_3=0$ (\ref{eq:E6L2L3ad}) give two scalar additive Lax equations $L_1=0$ and $L_2=0$ expressed by
\begin{equation}\label{eq:E6L1L2ad}
\begin{array}{l}
L_1(x)=\Big[\dfrac{(g+a_2)(g+b_2)(g+1)(g-a_0-b_0)}{(f-g)(x-g-1)}-(g-a_1+b_2-a_0)(g+a_2-b_1-b_0)\Big]y(x)
\\[5mm]\phantom{L_1(x)}
-\dfrac{x\prod_{i=1}^2(x+b_i-1)}{x-f-1}\Big[y(x-1)-\dfrac{(x-a_0-b_0-1)(x+a_2-1)}{(x-g-1)(x+b_1-1)}y(x)\Big]
\\[5mm]\phantom{L_1(x)}
-\dfrac{(x-a_0-b_0)\prod_{i=1}^2(x+a_i)}{x-f}\Big[y(x+1)-\dfrac{(x-g)(x+b_1)}{(x-a_0-b_0)(x+a_2)}y(x)\Big],
\\[5mm]
L_2(x)=C_0(f-x)\o{y}(x)-(x-a_0-b_0)(x+a_2)y(x+1)+(x+b_1)(x-g)y(x).
\end{array}
\end{equation}

The additive Lax form  of scalar type $L_1=0$ and $L_2=0$ (\ref{eq:E6L1L2ad}) is equivalent to the scalar ones in \cite{KNY17,  Nagao16} by using suitable gauge transformations of $y(x)$. On the other hand, the differential $3 \times 3$ matrix Lax form has been given as a certain Fuchsian system of differential equations in \cite{Boalch09}, and in \cite{AB06} a certain additive Lax form of $2 \times 2$ matrix type has been constructed utilizing moduli spaces of difference connections on $\mathbb{P}^1$ for type $d$-$E_6^{(1)}$, called the difference Painlev\'e {\rm VI} there.

\noindent
%E6ad
{\bf(e)} Determinant formulae of hypergeometric special solutions

The hypergeometric solutions are constructed as the explicit forms
\begin{equation}\label{eq:E6solad}
\begin{array}{l}
\dfrac{f+a_i}{f+b_j}=\alpha\dfrac{T_{a_i}(\tau_{m,n})T_{a_i}^{-1}(\tau_{m+1,n-1})}
{T_{b_j}^{-1}(\tau_{m,n})T_{b_j}(\tau_{m+1,n-1})} \quad  (i, j=1,2),\quad 
g=-a_2+\beta\dfrac{T_{a_2}(\o{\tau}_{m,n})T_{a_2}^{-1}(\tau_{m+1,n-1})}
{T_{a_1}(\tau_{m,n})T_{a_1}(\o{\tau}_{m+1,n-1})},
\end{array}
\end{equation}
where the determinant $\tau_{m,n}$ is given by
\begin{equation}\label{eq:E6tauad}
\begin{array}l
\ds \tau_{m,n}=\det\Big[(b_1)_i(a_1-j)_j{}_3F_2 \Big(\substack{\displaystyle{b_1+i, b_2,-(m+n)}\\[3mm]{\displaystyle{a_1-j, a_2}}},1\Big)\Big]^n _{i,j=0}, \\[5mm]
\alpha=-\dfrac{(a_i+m+n)(a_i-1)^n(b_j-1)^n}{a_i^{n+1}b_j^n}\dfrac{\prod_{k=1}^2(b_k-a_i)}{\prod_{k=1}^2(a_k-b_j)},\\[5mm]
\beta=-\dfrac{b_1(a_2+m+n)(a_2-1)^n(b_2-a_2)}{a_2^{n+1}(b_1-a_1)}.
\end{array}
\end{equation}
  
These determinant formulae of the generalized hypergeometric solutions (\ref{eq:E6solad}) are given in terms of the hypergeometric function ${}_3F_2$. A certain $1 \times 1$ determinant formula has been constructed in terms of the hypergeometric function ${}_3F_2$ in \cite{Kajiwara08, KNY17}.

%%%%%%%%%%%%%%%%%%%%%%%%
\subsection{Case $d$-$D_4^{(1)}$}\label{subsec:D4ad}

%D4ad
\noindent
{\bf(a)} Setting of the Pad\'e interpolation problem on the additive grid

In Table (\ref{eq:Ylistad}) the interpolated function and the interpolated sequence are set up as
\begin{equation}\label{eq:D4Yad}
Y(x):=\displaystyle c^x\dfrac{\Gamma(a_1)\Gamma(x+b_1)}{\Gamma(x+a_1)\Gamma(b_1)},\quad Y_s=\displaystyle c^s\dfrac{(b_1)_s}{(a_1)_s},
\end{equation}
and in Table (\ref{eq:Tlistad}) the time evolution is chosen as
\begin{equation}\label{eq:D4Tad}
T: (a_1,b_1,c,m,n) \mapsto (a_1+1,b_1+1,c,m,n).
\end{equation}

%D4ad
\noindent
{\bf(b)} Contiguity three term relations

By Definition (\ref{eq:GKHad}) we have the basic quantities
\begin{equation}\label{eq:D4GKHad}
\displaystyle G(x)=c\dfrac{x+b_1}{x+a_1},\quad 
K(x)=\dfrac{a_1(x+b_1)}{b_1(x+a_1)},\quad H(x)=b_1(x+a_1),
\end{equation}
and by the expression (\ref{eq:Drelationad}) we obtain the Casorati determinants
\begin{equation}\label{eq:D4Dad}
\begin{array}{l}
D_1(x)=\sfrac{c_0(x-f){\mathcal N}(x)Y(x)}{G(x)_{\rm den}},\quad
D_2(x)=\sfrac{c_1(x-m-n){\mathcal N}(x)Y(x)}{K(x)_{\rm den}},
\\[5mm]
D_3(x)=\sfrac{c_2{\mathcal N}(x)Y(x)(x+b_1)}{H(x)},
\end{array}
\end{equation}
where $f,c_0,c_1$ and $c_2$ are constants depending on parameters $a_1, b_1, c \in \C^{\times}$ and $m,n \in \Z_{\geq 0}$ but independent of $x$. Then the contiguity relations $L_2=0$ and $L_3=0$ are expressed by
\begin{equation}\label{eq:D4L2L3ad}
\begin{array}{l}
L_2(x)=C_0(f-x)\o{y}(x)-(x-m-n)y(x+1)+\sfrac{(x+b_1)y(x)}{g},\\
L_3(x)=C_1(x-\o{f}-1)y(x)+\sfrac{(x+a_1)\o{y}(x)}{g}
-cx\o{y}(x-1),
\end{array}
\end{equation}
where $C_0=\sfrac{b_1c_0}{c_1}$, $C_1=\sfrac{a_1\o{c}_0}{c_1}$ and $g=\sfrac{c_1}{c_2}$.

Take note that in the items (c) and (d) below we study the contiguity relations $L_2=0$ and $L_3=0$ (\ref{eq:D4L2L3ad}) for generic complex parameters $a_0, b_0$ (replacing $m, n \in \Z_{\geq 0}$ by $a_0, b_0 \in \C^{\times}$) and generic variables $f, g \in \P^1$ (depending on parameters $a_i, b_i, c \in \C^{\times}, i=0,1$) apart from the Pad\'e interpolation problem (\ref{eq:padead}) with eqs. (\ref{eq:D4Yad}) and (\ref{eq:D4Tad}) (see Remark \ref{rem:genericad}).

%D4ad
\noindent
{\bf(c)} The time evolution equations of the $d$-Painlev\'e equation

Compatibility of the contiguity relations $L_2=0$ and $L_3=0$ (\ref{eq:D4L2L3ad}) gives the evolution equations and the product $C_0C_1$ as follows.
\begin{equation}\label{eq:D4eqad}
\begin{array}{l}
g\u{g}=\dfrac{c(f+1)(f-a_0-b_0)}{(f+a_1)(f+b_1)},\quad f+\o{f}=\dfrac{a_1+a_0}{cg-1}+\dfrac{b_1+b_0}{g-1}+a_0+b_0-1
\end{array}
\end{equation}
and
\begin{equation}\label{eq:D4C0C1ad}
C_0C_1=\sfrac{(g-1)(cg-1)}{g^2}.
\end{equation}

The evolution equations (\ref{eq:D4eqad}) are equivalent to the $d$-Painlev\'e equation of type $D_4^{(1)}$ given in \cite{Kajiwara08, KNY17, Nagao16, RGTT01, Sakai07}. The eight singular points in coordinates $(f,g)$ are on the four lines $f=0$, $g=0$, $f=\infty$ and $g=\infty$ as follows.
\begin{equation}\label{eq:D48pad}
\begin{array}{l}
(f,g)=(-a_1,\infty), (-b_1,\infty), (\sfrac{1}{\varepsilon},\sfrac{\{1+(a_0+a_1)\varepsilon\}}{c})_2,\\
\phantom{(f,g)=} (\sfrac{1}{\varepsilon}, 1+(b_0+b_1)\varepsilon)_2, (-1,\infty), (a_0+b_0, \infty).
\end{array}
\end{equation}

Here, the third point is a double point at $(\infty,\sfrac{1}{c})$ with the gradient $f(g-\sfrac{1}{c})=\sfrac{a_0+a_1}{c}$ and the fourth point is a double point at $(\infty,1)$ with the gradient $f(g-1)=b_0+b_1$. (The meaning of the two double points is also written in \cite{KNY17}.)

%D4ad
\noindent
{\bf(d)} The additive difference Lax form of scalar type

The contiguity relations $L_2=0$ and $L_3=0$ (\ref{eq:D4L2L3ad}) give two scalar additive Lax equations $L_1=0$ and $L_2=0$ expressed by
\begin{equation}\label{eq:D4L1L2ad}
\begin{array}{l}
L_1(x)=(g-1)(c g-1)\Big[\dfrac{c(a_1+a_0)}{c g-1}+\dfrac{b_1+b_0}{g-1}-\dfrac{x+f+a_1+b_1}{g}\Big]y(x)
\\[5mm]\phantom{L_1(x)}
-\dfrac{cx(x+b_1-1)}{x-f-1}\Big[y(x-1)-\dfrac{g(x-a_0-b_0-1)}{x+b_1-1}y(x)\Big]
\\[5mm]\phantom{L_1(x)}
-\dfrac{(x-a_0-b_0)(x+a_1)}{x-f}\Big[y(x+1)-\dfrac{x+b_1}{g(x-a_0-b_0)}y(x)\Big],
\\[5mm]
L_2(x)=(f-x)\o{y}(x)-(x-a_0-b_0)y(x+1)+\sfrac{(x+b_1)y(x)}{g}.
\end{array}
\end{equation}

The additive Lax form  of scalar type  (\ref{eq:D4L1L2ad}) is equivalent to the scalar ones in \cite{KNY17,  Nagao16} by using suitable gauge transformations of $y(x)$. On the other hand, in \cite{AB06} a certain additive Lax form of 2 $\times$ 2 matrix type has been obtained utilizing moduli spaces of difference connections on $\mathbb{P}^1$ for type $d$-$D_4^{(1)}$, called the difference Painlev\'e {\rm V}. Concerning the differential Lax form for type $d$-$D_4^{(1)}$, the $2 \times 2$ matrix Lax pair and the scalar one have been derived respectively in \cite{GORS98} and \cite{KNY17} by using a Schlesinger transformation of differential equations.

\noindent
%D4ad
{\bf(e)} Determinant formulae of hypergeometric special solutions

The hypergeometric solutions are constructed as the explicit forms
\begin{equation}\label{eq:D4solad}
\begin{array}{l}
\dfrac{f+a_1}{f+b_1}=\alpha\dfrac{T_{a_1}(\tau_{m,n})T_{a_1}^{-1}(\tau_{m+1,n-1})}
{T_{b_1}^{-1}(\tau_{m,n})T_{b_1}(\tau_{m+1,n-1})},\quad 
\dfrac{1}{g}=1-\dfrac{b_1(c-1)}{b_1-a_1}
\dfrac{\o{\tau}_{m,n}\tau_{m+1,n-1}}
{T_{a_1}(\tau_{m,n})T_{a_1}(\o{\tau}_{m+1,n-1})},
\end{array}
\end{equation}
where the determinant $\tau_{m,n}$ is given by
\begin{equation}\label{eq:D4tauad}
\begin{array}l
\ds \tau_{m,n}=\det\Big[(b_1)_i(a_1-j)_j{}_2F_1 \Big(\substack{\displaystyle{b_1+i,-(m+n)}\\[3mm]{\displaystyle{a_1-j}}},c\Big)\Big]^n _{i,j=0}, \quad 
\alpha=c\dfrac{(a_1+m+n)(a_1-1)^n(b_1-1)^n}{a_1^{n+1}b_1^n}.
\end{array}
\end{equation}
  
These determinant formulae of the Gauss hypergeometric solutions (\ref{eq:D4solad}) are given in terms of the 
%\red{terminating}\footnote{\red{See Remark \ref{rem:terminating}. The solution solutions (\ref{eq:D4solad}) are expected to be given in terms of the non-terminating ${}_2F_1$, replacing $m\in\Z_{\ge 0}$ by $a_0\in\C^{\times}$ apart from Pad\'e interpolation problem.}} 
hypergeometric function ${}_2F_1$. A certain $1 \times 1$ determinant formula has been constructed in terms of the 
%\red{non-terminating}\footnote{\red{The general determinant formulae have not  been given yet in terms of the non-terminating ${}_2F_1$ as far as we know.}} 
hypergeometric function ${}_2F_1$ (e.g. \cite{KNY17}).
 
%%%%%%%%%%%%%%%%%%%%%%%%
\subsection{Case $d$-$A_3^{(1)}$}\label{subsec:A3ad}

%A3ad
\noindent
{\bf(a)} Setting of the Pad\'e interpolation problem on the additive grid

In Table (\ref{eq:Ylistad}) the interpolated function and the interpolated sequence are set up as
\begin{equation}\label{eq:A3Yad}
Y(x):=d^x\dfrac{\Gamma(x+b_1)}{\Gamma(b_1)},\quad Y_s=d^s(b_1)_s,
\end{equation}
and in Table (\ref{eq:Tlistad}) the time evolution is chosen as
\begin{equation}\label{eq:A3Tad}
T: (b_1,d,m,n) \mapsto (b_1+1,d,m,n).
\end{equation}\\

%A3ad
\noindent
{\bf(b)} Contiguity three term relations

By Definition (\ref{eq:GKHad}) we have the basic quantities
\begin{equation}\label{eq:A3GKHad}
\displaystyle G(x)=d(x+b_1), \quad 
K(x)=\dfrac{x+b_1}{b_1},\quad H(x)=b_1,
\end{equation}
and by the expression (\ref{eq:Drelationad}) we obtain the Casorati determinants
\begin{equation}\label{eq:A3Dad}
\begin{array}{l}
D_1(x)=\sfrac{c_0(x-f){\mathcal N}(x)Y(x)}{G(x)_{\rm den}},\quad
D_2(x)=\sfrac{c_1(x-m-n){\mathcal N}(x)Y(x)}{K(x)_{\rm den}},
\\[5mm]
D_3(x)=\sfrac{c_2{\mathcal N}(x)Y(x)(x+b_1)}{H(x)},
\end{array}
\end{equation}
where $f,c_0,c_1$ and $c_2$ are constants depending on parameters $b_1, d \in \C^{\times}$ and $m,n \in \Z_{\geq 0}$ but independent of $x$. Then the contiguity relations $L_2=0$ and $L_3=0$ are expressed by
\begin{equation}\label{eq:A3L2L3ad}
\begin{array}{l}
L_2(x)=C_0(f-x)\o{y}(x)-(x-m-n)y(x+1)+\sfrac{(x+b_1)y(x)}{g},\\
L_3(x)=C_1(x-\o{f}-1)y(x)+\sfrac{\o{y}(x)}{g}-dx\o{y}(x-1),
\end{array}
\end{equation}
where $C_0=\sfrac{b_1c_0}{c_1}$, $C_1=\sfrac{\o{c}_0}{c_1}$ and $g=\sfrac{c_1}{c_2}$.\\

Take note that in the items (c) and (d) below we study the contiguity relations $L_2=0$ and $L_3=0$ (\ref{eq:A3L2L3ad}) for generic complex parameters $a_0, b_0$ (replacing $m, n \in \Z_{\geq 0}$ by $a_0, b_0 \in \C^{\times}$) and generic variables $f, g \in \P^1$ (depending on parameters $a_0, b_0, b_1, d \in \C^{\times}$) apart from the Pad\'e interpolation problem (\ref{eq:padead}) with eqs. (\ref{eq:A3Yad}) and (\ref{eq:A3Tad}) (see Remark \ref{rem:genericad})\\

%A3ad
\noindent
{\bf(c)} The time evolution equations of the $d$-Painlev\'e equation

Compatibility of the contiguity relations $L_2=0$ and $L_3=0$ (\ref{eq:A3L2L3ad}) gives the evolution equations and the product $C_0C_1$ as follows.
\begin{equation}\label{eq:A3eqad}
\begin{array}{l}
g\u{g}=\dfrac{f+b_1}{d(f+1)(f-a_0-b_0)},\quad f+\o{f}=\dfrac{1}{dg}+\dfrac{b_0+b_1}{g-1}+a_0+b_0-1
\end{array}
\end{equation}
and
\begin{equation}\label{eq:A3C0C1ad}
C_0C_1=\sfrac{d(g-1)}{g}.
\end{equation}

The evolution equations (\ref{eq:A3eqad}) are equivalent to the $d$-Painlev\'e equation of type $A_3^{(1)}$ given in \cite{Kajiwara08, KNY17, Nagao16, RGTT01, Sakai07}. The eight singular points in coordinates $(f,g)$ are on the four lines $f=0$, $g=0$, $f=\infty$ and $g=\infty$ as follows.
\begin{equation}\label{eq:A38pad}
\begin{array}{l}
(f,g)=(-1,0), (a_0+b_0, 0), (\sfrac{1}{\varepsilon},\sfrac{\varepsilon}{d}(1+m\varepsilon))_3, (\sfrac{1}{\varepsilon}, 1+(b_0+b_1)\varepsilon)_2, (-b_1, \infty).
\end{array}
\end{equation}

Here, the third point is a triple point at $(\infty,0)$ and the fourth point is a double point at $(\infty,1)$ with the gradient $f(g-1)=b_0+b_1$. (The meaning of the triple and double points is also written in \cite{KNY17}.)

%A3ad
\noindent
{\bf(d)} The additive difference Lax form of scalar type

The contiguity relations $L_2=0$ and $L_3=0$ (\ref{eq:A3L2L3ad}) give two scalar additive Lax equations $L_1=0$ and $L_2=0$ expressed by
\begin{equation}\label{eq:A3L1L2ad}
\begin{array}{l}
L_1(x)=(g-1)\Big[\dfrac{1}{dg}+\dfrac{b_1+b_0}{g-1}-(x+f-a_0-b_0)\Big]y(x)
\\[5mm]\phantom{L_1(x)}
-\dfrac{x(x+b_1-1)}{x-f-1}\Big[y(x-1)-\dfrac{g(x-a_0-b_0-1)}{x+b_1-1}y(x)\Big]
\\[5mm]\phantom{L_1(x)}
-\dfrac{x-a_0-b_0}{d(x-f)}\Big[y(x+1)-\dfrac{x+b_1}{g(x-a_0-b_0)}y(x)\Big],
\\[5mm]
L_2(x)=(f-x)\o{y}(x)-(x-a_0-b_0)y(x+1)+\sfrac{(x+b_1)y(x)}{g}.\\
\end{array}
\end{equation}

The additive Lax form of scalar type $L_1=0$ and $L_2=0$ (\ref{eq:A3L1L2ad}) is equivalent to the scalar ones in \cite{KNY17,  Nagao16} by using suitable gauge transformations of $y(x)$. Concerning the differential Lax form for type $d$-$A_3^{(1)}$, the $2 \times 2$ matrix Lax pair and the scalar one have been derived respectively in \cite{GORS98} and \cite{KNY17} by using a Schlesinger transformation of differential equations.

\noindent
%A3ad
{\bf(e)} Determinant formulae of hypergeometric special solutions

The hypergeometric solutions are constructed as the explicit forms
\begin{equation}\label{eq:A3solad}
\begin{array}{l}
f=-b_1+\dfrac{db_1^n}{d(b_1-1)^n}\dfrac{T_{b_1}^{-1}(\tau_{m,n})T_{b_1}(\tau_{m+1,n-1})}{\tau_{m,n}\tau_{m+1,n-1}}
,\quad 
g=1-db_1\dfrac{\o{\tau}_{m,n}\tau_{m+1,n-1}}
{\tau_{m,n}\o{\tau}_{m+1,n-1}},
\end{array}
\end{equation}
where the determinant $\tau_{m,n}$ is given by
\begin{equation}\label{eq:A3tauad}
\begin{array}l
\ds \tau_{m,n}=\det\Big[(b_1)_i(-(m+n))_j {}_2F_0 \Big(\substack{\displaystyle{b_1+i,-(m+n)+j}\\[3mm]{\displaystyle{0}}},d\Big)\Big]^n _{i,j=0}.
\end{array}
\end{equation}
  
These determinant formulae of the Kummer hypergeometric solutions (\ref{eq:A3solad}) are given in terms of the hypergeometric function ${}_2F_0$. A certain $1 \times 1$ determinant formula has been constructed in terms of the hypergeometric function ${}_1F_1$ (e.g. \cite{KNY17}). The terminating ${}_1F_1$ can be transformed into the terminating ${}_2F_0$ (see Remark \ref{rem:trans}).
%%%%%%%%%%%%%%%%%%%%%%%%

\section{Conclusions}\label{sec:conc}
\subsection{Summary}

\noindent
In this paper for the interpolated function $Y(x)$ and the interpolated sequence $Y_s$ given in Table \ref{tb:Ylistad} of Section \ref{subsec:itema}, we set up the Pad\'e interpolation problem on the additive gird, related to the $d$-Painlev\'e equations of type $E_7^{(1)}$, $E_6^{(1)}$, $D_4^{(1)}$ and $A_3^{(1)}$. Then for the time evolution $T$ given in Table \ref{tb:Tlistad} of Section \ref{subsec:itema}, we set up another Pad\'e interpolation problem on the additive gird. By choosing these suitable problems, we derived the evolution equations, the Lax pairs of scalar type and the determinant formulae of the special solutions for the corresponding $d$-Painlev\'e equations. The main results were given in Section \ref{sec:main results}. 

\subsection{Problems}

\noindent
As is shown in Table \ref{tb:work} of Section \ref{subsec:pade method}, some open problems related to the results of this paper are as follows:

{\bf 1.} One may be interested in studying whether the Pad\'e interpolation method on the additive quadratic grid can be applied to the additive difference ($d$-) Painlev\'e equations. 

{\bf 2.} Differently from the additive grid, it may be interesting to investigate whether the Pad\'e method on the differential grid (i.e. Pad\'e approximation) can be also applied to the $d$-Painlev\'e equations by using a Schlesinger transformation of linear differential equations.

{\bf 3.} It may be interesting to study whether the Pad\'e method can be further applied to other generalized $d$-Painlev\'e systems, for example an additive difference analogue of the Garnier system and a higher order Painlev\'e system, which are called ``{\it $d$-Garnier system}" \cite{DST13, DT14, OR16} and ``{\it higher order $d$-Painlev\'e system}'', respectively.\\

\section*{Acknowledgments}

\noindent
The author is grateful to Professor Yasuhiko Yamada for valuable discussions on this research. He also thanks Professor Kenji Kajiwara for stimulating comments. This work was partially supported by JSPS KAKENHI (19K14579) and Expenses Revitalizing Education and Research of Akashi College.

%%%%%%%%%%%
\appendix
\section{Sufficiency for the compatibility of the Lax pair}\label{sec:Proof}

In Section \ref{subsec:E7ad} we gave the $d$-$E_7^{(1)}$ equation (\ref{eq:E7eqad}) as the necessary condition for the compatibility of the Lax pair (\ref{eq:E7L1L2ad}). In this appendix, we prove that the $d$-$E_7^{(1)}$ equation is the sufficient condition for the compatibility of the Lax pair. 

As in Figure \ref{fig:L1}, eliminating $\o{y}(x)$ and $\o{y}(x-1)$ from $L_2(x)= L_2(x-1)=L_3(x)=0$ (\ref{eq:E7L2L3ad}), one constructs the linear equation $L_1=0$ among 
$y(x+1)$, $y(x)$ and $y(x-1)$, where
\begin{equation}\label{eq:E7L1ad}
\begin{array}l
\ds L_1(x)= \frac{(x-a_0-b_0)\prod_{i=1}^3(x+a_i)}{x-f}y(x+1)+\frac{x\prod_{i=1}^3(x+b_i-1)}{(x-f-1)}y(x-1)\\[5mm]
\ds\phantom{L_1(x): }-\frac{1}{x-h}\Big[\frac{A_2(x)(x-g)}{x-f}+\frac{V(x-1)}{(x-f-1)(x-g-1)}\Big]y(x)
\end{array}
\end{equation}
and
\begin{equation}\label{eq:E7Vad}
V(x)=(x-h)(x-h+1)A_1(x)-C_0C_1(x-f)(x-\o{f}).
\end{equation}
Here, the variable $\o{f}$ and the product $C_0C_1$ in (\ref{eq:E7Vad}) should be viewed as functions in terms of $f$ and $g$, and they are determined in (\ref{eq:E7eqad}) and (\ref{eq:E7C0C1ad}),  respectively. The expression $L_1$ (\ref{eq:E7L1ad}) is rewritten into (\ref{eq:E7L1L2ad}) by using (\ref{eq:E7eqad}) and (\ref{eq:E7C0C1ad}). 

\begin{lem}\label{lem:E7L1p}
The expression $(x-f)(x-f-1)L_1(x)$ (\ref{eq:E7L1ad}) (or (\ref{eq:E7L1L2ad})) has the following characterization:\\
\hspace{5mm}(i) It is a linear equation among $y(x+1)$, $y(x)$ and $y(x-1)$, and the coefficients of these terms are polynomials of degree $5$ in $x$.\\
\hspace{5mm}(ii) The coefficients of $y(x+1)$ (resp. $y(x-1)$) have zeros at $x=-a_1$,$-a_2$,$-a_3$, $a_0+b_0$ (resp. $x=-b_1+1$,$-b_2+1$, $-b_3+1$, $0$).\\ 
\hspace{5mm}(iii) Under the conditions 
\begin{equation}\label{eq:cond}
\begin{array}l
\displaystyle\frac{y(x+1)}{y(x)}=1+\frac{a_0}{x}+\frac{a_0(a_0-1)/2}{x^2}+\frac{w}{x^3}+O\Big(\frac{1}{x^4}\Big), \\[5mm]
\displaystyle \frac{y(x-1)}{y(x)}=1-\frac{a_0}{x}+\frac{a_0(a_0-1)/2}{x^2}-\frac{w}{x^3}+O\Big(\frac{1}{x^4}\Big),
\end{array}
\end{equation}
the terms $x^5, \ldots, x^2$ in the expression $(x-f)(x-f-1)L_1(x)$ vanish, namely $(x-f)(x-f-1)L_1(x)=O(x^1)$ around $x=\infty$. Here, $w \in \C$ is an arbitrary constant.\\
\hspace{5mm}(iv) The equation $(x-f)(x-f-1)L_1=0$ holds at the two points $x=f, f+1$, where
\begin{equation}\label{eq:y}
\dfrac{y(f+1)}{y(f)} =\dfrac{(f+b_1)(f+b_3)(f-g)}{(f+a_2)(f-a_0-b_0)(f-h)}.
\end{equation}
Conversely, the expression $(x-f)(x-f-1)L_1(x)$ is uniquely characterized by these properties $(i)-(iv)$.
\sq
\end{lem}

\prf
The property (i) is obtained by the relations (\ref{eq:E7C0C1ad}). Concretely, the expression $\frac{V(x-1)}{x-g-1}$ reduces to a polynomial of degree 5 in $x$ under the first relation of (\ref{eq:E7C0C1ad}). Moreover, the coefficient of the term $y(x)$ is obtained as a polynomial of degree 5 in $x$ by using the second relation of (\ref{eq:E7C0C1ad}). The property (ii) is trivial. The property (iii) can easily be checked by the condition (\ref{eq:cond}). The property (iv) follows by substituting $x=f, f+1$ into the equation $L_1(x)=0$. 
\qed

\begin{rem}
Two points $x=f, f+1$ are apparent singularities in the sense that at those two points the equation $(x-f)(x-f-1)L_1(x)=0$ (\ref{eq:E7L1ad}) is satisfied under the same condition (in this case (\ref{eq:y})). \sq
\end{rem}

Similarly, as in Figure \ref{fig:L1*}, eliminating $y(x)$ and $y(x+1)$ from $L_2(x)= L_3(x)=L_3(x+1)=0$ (\ref{eq:E7L2L3ad}), we obtain the linear equation $L_1^*=0$ among
$\o{y}(x+1)$, $\o{y}(x)$ and $\o{y}(x-1)$, where
\begin{equation}\label{eq:E7L1*ad}
\begin{array}l
\ds L_1^*(x)= \frac{(x-a_0-b_0)(x+a_1+1)(x+a_2)(x+a_3+1)}{x-\o{f}}\o{y}(x+1)
\\[5mm]\phantom{L_1^*(x)}
\ds+\frac{x(x+b_1)(x+b_2-1)(x+b_3)}{(x-\o{f}-1)}\o{y}(x-1)
\ds-\frac{1}{x-h}\Big[\frac{A_2(x)(x-g-1)}{x-\o{f}-1}+\frac{V(x)}{(x-\o{f})(x-g)}\Big]y(x).
\end{array}
\end{equation}

\begin{figure}[ht]
\begin{center}\setlength{\unitlength}{1mm}
\begin{picture}(50,30)(-5,-5)
\put(0,22){\line(1,0){43}}
\put(-2,25){$\o{y}(x-1)$}
\put(18,25){$\o{y}(x)$}
\put(39,25){$\o{y}(x+1)$}
\put(18,-5){$y(x)$} 
\put(39,-5){$y(x+1)$}
\put(23,21){\line(1,0){20}}
\put(43,1){\line(0,1){20}}
\put(23,21){\line(1,-1){20}}
\put(1,21){\line(1,0){20}}
\put(21,1){\line(0,1){20}}
\put(1,21){\line(1,-1){20}}
\put(31,14){$L_3(x+1)$}
\put(10,14){$L_3(x)$}
\put(22,0){\line(1,0){20}}
\put(22,0){\line(0,1){20}}
\put(22,20){\line(1,-1){20}}
\put(23,4){$L_2(x+1)$}
\put(-15,25){$L_1^*(x)$:}
\end{picture}
\caption{Derivation of $L_1^*(x)$}\label{fig:L1*}
\end{center}
\end{figure}

The following Lemma (and its proof) is similar to Lemma \ref{lem:E7L1p}.
\begin{lem}\label{lem:E7L1*p}
The expression $(x-\o{f})(x-\o{f}-1)L_1^*(x)$ (\ref{eq:E7L1*ad}) has the following characterization:\\
\hspace{5mm}(i) It is a linear three term expression among $\o{y}(x+1)$ and $\o{y}(x)$ and $\o{y}(x-1)$, and the coefficients of these terms are polynomials of degree $5$ in $x$.\\
\hspace{5mm}(ii) The coefficients of $\o{y}(x+1)$ (resp. $\o{y}(x-1)$) have zeros at $x=-a_1-1$,$-a_2$,$-a_3-1$, $a_0+b_0$ (resp. $x=-b_1$,$-b_2+1$, $-b_3$, $0$).\\ 
\hspace{5mm}(iii) Under the conditions
\begin{equation}\label{eq:cond*}
\begin{array}l
\displaystyle\frac{\o{y}(x+1)}{\o{y}(x)}=1+\frac{a_0}{x}+\frac{\o{a}_0(a_0-1)/2}{x^2}+\frac{\o{w}}{x^3}+O\Big(\frac{1}{x^4}\Big), \\[5mm]
\displaystyle \frac{\o{y}(x-1)}{\o{y}(x)}=1-\frac{a_0}{x}+\frac{a_0(a_0-1)/2}{x^2}-\frac{\o{w}}{x^3}+O\Big(\frac{1}{x^4}\Big),
\end{array}
\end{equation}
the terms $x^5, \ldots, x^2$ in the expression $(x-\o{f})(x-\o{f}-1)L_1^*(x)$ vanish, namely $(x-\o{f})(x-\o{f}-1)L_1^*(x)=O(x^1)$ around $x=\infty$. Here, $w \in \C$ is the same arbitrary constant as in (\ref{eq:cond}).\\
\hspace{5mm}(iv) The equation $(x-\o{f})(x-\o{f}-1)L_1^*=0$ holds at the two points $x=\o{f}, \o{f}+1$ where
\begin{equation}\label{eq:yu}
\dfrac{\o{y}(\o{f}+1)}{\o{y}(\o{f})} =\dfrac{(\o{f}+b_2)(\o{f}+1)(\o{f}+h-1)}{(\o{f}+a_1+1)(\o{f}+a_3+1)(\o{f}-g)}.
\end{equation}
Conversely, the expression $(x-\o{f})(x-\o{f}-1)L_1^*(x)$ is uniquely characterized by these properties $(i)-(iv)$.
\sq
\end{lem}

The sufficiency for the compatibility means that $T(L_1(x))\propto L_1^*(x)$ holds when the $d$-$E_7^{(1)}$ equation (\ref{eq:E7eqad}) is satisfied. In order to prove the sufficiency, we characterize $L_1$ and $L_1^*$ as polynomials in terms of $x$, and compare these characterizations.

\begin{prop}\label{prop:Lax2}
The linear equations $L_1=0$ and $L_2=0$ (\ref{eq:E7L1L2ad}) for the  unknown function $y(x)$ are compatible
if and only if the $d$-$E_7^{(1)}$ equation (\ref{eq:E7eqad}) is satisfied. 
\sq
\end{prop}

\prf
The compatibility means that the shift operator $T$ changes the equation $L_1=0$ into the equation $L_1^{\ast}=0$, i.e. the commutativity in Figure \ref{fig:com}.
\begin{figure}[ht]
\begin{equation}\label{eq:classification2}\nonumber
\begin{array}{cccc}
L_1^* =0 \hspace{2mm}(\mbox{Lemma}\hspace{2mm}\ref{lem:E7L1*p}) &\Leftrightarrow&L_1^*=0\hspace{2mm}(\ref{eq:E7L1*ad})&\\
&&\uparrow\\
\uparrow \mbox{$T$-shift}\hspace{2mm}(\ref{eq:E7Tad})&&L_2 =L_3=0 \hspace{2mm}(\ref{eq:E7L2L3ad})&\\
&&\downarrow\\
L_1=0 \hspace{2mm}(\mbox{Lemma}\hspace{2mm}\ref{lem:E7L1p})\ &\Leftrightarrow& L_1=0\hspace{2mm}(\ref{eq:E7L1ad})&\Leftrightarrow L_1=0 \ (\ref{eq:E7L1L2ad}).
\end{array}
\end{equation}
\caption{Compatibility of $L_1(x)$ and $L_1^*(x)$}\label{fig:com}
\end{figure}

This commutativity is almost clear from the characterizations (i), (ii) of the equation $L_1=0$ (respectively $L_1^*=0$) in Lemma \ref{lem:E7L1p} (respectively Lemma \ref{lem:E7L1*p}).
The remaining task is to check that the operator $T$ changes expression (\ref{eq:y}) into expression (\ref{eq:yu}), utilizing the characterization (iii) of the equation $L_1=0$ (respectively $L_1^*=0$) and the first part of equation (\ref{eq:E7eqad}).
\qed

As the point of the proof, the following two are applied to type $d$-$E_7^{(1)}$ together: The first is that the equation $L_1(f,\o{f}, g)=0$ in terms of $f$, $\o{f}$ and $g$ is derived from the equations $L_2(f,g)=0$ and $L_3(\o{f},g)=0$ (see \cite{KNY17,NTY13,Yamada09-2,Yamada11}). The second is that the equation $L_1(f,\o{f}, g)=0$ is characterized as a polynomial in terms of $x$ (see \cite{Nagao16,NY18-1}).

%%%%%%%%%%%%%%%%%%%%%%

\end{document}